\documentclass[12pt]{article}

\usepackage{hyperref}

\usepackage{amsmath} 
\usepackage{amssymb}
\usepackage{amsfonts}
\usepackage{amsthm}
\usepackage{graphicx}
\usepackage{bm}

\newtheorem{theorem}{Theorem}
\newtheorem{lemma}{Lemma}
\newtheorem{corollary}{Corollary}
\newtheorem{proposition}{Proposition}

\theoremstyle{definition}

\newtheorem{definition}{Definition}

\begin{document}
	\begin{center}
		\Large
	\textbf{Irreversible quantum evolution with quadratic generator: Review}\footnote{This work is supported by the Russian Science Foundation under grant~19-11-00320.}	
	
		\large 
		\textbf{A.E. Teretenkov}\footnote{Steklov Mathematical Institute of Russian Academy of Sciences,
			ul. Gubkina 8, Moscow 119991, Russia\\ E-mail:\href{mailto:taemsu@mail.ru}{taemsu@mail.ru}}
		\end{center}
		
			\footnotesize
			We review results on GKSL-type equations with multi--modal generators which are quadratic in bosonic or fermionic creation and annihilation operators. General forms of such equations are presented. The Gaussian solutions are obtained in terms of equations for the first and the second moments. Different approaches for their solutions are discussed.
			\normalsize

	\section{Introduction}
	
Generators of quantum Markov semi--groups which are quadratic in bosonic or fermionic operators 
for both unitary and irreversible evolution are a starting point for other considerations both exact
and approximate. Thus, it is useful to review main results on them and collect main formulae in one
concise text and in the same notation to simplify their application for the readers. 
We consider GKSL generators, but unitary evolutions could be considered as a particular case. 
We consider the case with a finite, but arbitrary number of bosonic or fermionic modes because of its applications in quantum optics and quantum information.

This work is organized as follows. We separate an historical and literature review in
Sec.~\ref{sec:HistLitRew} in such a way that if the reader is not interested in it and just wants 
to find a specific formula, then he could freely skip this section. Then the bosonic and fermionic
case are considered in Sec.~\ref{sec:bosonic} and \ref{sec:femionic}, respectively. These sections are also mostly independent. Nevertheless, Sec.~\ref{sec:bosonic} and \ref{sec:femionic} are organized 
in similar subsections and could be read in parallel.

Subsections ~\ref{subsec:basicBos} and \ref{subsec:basicFerm} introduce the basic formulae for
calculations with quadratic and linear forms in creation and annihilation operators. 
Such calculations are widely used and could be useful by themselves not only for the aims considered
here. Subsections ~\ref{subsec:QuadBos} and \ref{subsec:QuadFerm} consider superoperators, in
particular, \textit{Gorini-Kossakowski-Sudarshan-Lindblad} (GKSL) generators which also have only quadratic and linear terms. In Subsec.~\ref{subsec:GaussBos} and \ref{subsec:GaussFerm} we define 
the Gaussian states and consider several representations of them.

The main results are presented in Subsec.~\ref{subsec:DynBos} and \ref{subsec:DynFerm}, where the solutions of GKSL equations with quadratic generators are obtained in terms of equations for the second and the first moments. Several representations of these equations are discussed which lead to different approaches to their solutions.

In the final section we sum up the results presented in this article and suggest possible directions of the further study.

\section{Historical and literature review}\label{sec:HistLitRew}

Irreversible quadratic quantum evolutions are direct generalization of unitary dynamics with quadratic generators. Unitary dynamics with quadratic generators of general form were systematically studied by
K.\,O.~Friedrichs\cite{Fried1953} and F.\,A.~Berezin\cite{Ber86}. Further study of unitary dynamics
was developed in the works by I.\,A.~Malkin, V.\,I.~Man'ko and V.\,V.~Dodonov \cite{Manko79, Manko87, dodonov2003theory}.  
Some modern studies on unitary dynamics with quadratic Hamiltonians could be found in works by A.\,M.~Chebotraev and T.\,V.~Tlyachov \cite{Cheb12, Cheb11, chebotarev2014normal}. One of the main
areas of application of quadratic multi--modal Hamiltonians is quantum optics, especially in case of
parametric approximation (see Ref.~\cite{achmanov1964}, Ch.~II,  Ref.~\cite{scalli2003}, Ch.~16, or Ref.~\cite{maimistov2013nonlinear}, Sec.~7.1.2). Coherent states for multi--modal parametric
systems were considered in Ref.~\cite{dodonov1995even}. Modern applications could be found in works by A.\,S.~Chirkin et al \cite{chirkin2007statistic, Chirkin2007, chirkin2009four, saygin2010simultaneous, tlyachev2014canonical, tlyachev2013new}. Also quadratic Hamiltonians arise in opto--mechanical problems  \cite{huang2009entangling, huang2009squeezing}. Another source of such Hamiltonians is the approximate quantization method by N.\,N.~Bogolyubov \cite{Bogol1947}. (See also Ref. 80 for contemporary discussion.) Discussion of the unitary dynamics with non-stationary (time-dependent) Hamiltonians could be found in Refs.~\cite{chernikov1968system}--\cite{Bud1987One}. Physical applications of such Hamiltonians were discussed in Ref.~\cite{castanos2006squeezing}. Quantum evolution with a quadratic generator is also closely related to the classical one, which was studied in the general case by J.~Williamson \cite{williamson1936algebraic, williamson1940algebraic, williamson1937normal}. The quasi-classical approximation for squeezed states was discussed in Ref.~\cite{alekseev2009squeezed}. More detailed bibliography for bosonic unitary evolution could be found in Refs.~\cite{dodonov2003theory} and \cite{dodonov2002nonclassical}. The fermionic case without dissipation was considered in Ref.~\cite{Ber86} by F.\,A.~Berezin. The unified approach both for the bosonic reversible dynamics and the fermionic one was considered by V.\,I.~Man'ko and V.\,V.~Dodonov.\cite{DodMan83}

We consider reversible (non-unitary) evolution with quadratic generators. From the historical point of view one should mention Ref.~\cite{landau1927} by L.\,D.~Landau, which is famous for the first appearance of the concept of the density matrix. What is interesting for this review is that equations for irreversible quantum evolution of the density matrix were also introduced there. As it was mentioned on p. 183 in Ref.~\cite{dodonov2003theory} and in Sec. 23.1 of Ref.~\cite{perelomov1987} in the harmonic oscillator case one has (in modern notation)
\begin{equation*}
\dot{\rho}_t = \gamma \left(\hat{a} \rho_t \hat{a}^{\dagger} - \frac12 \hat{a}^{\dagger}\hat{a} \rho_t - \frac12 \rho_t \hat{a}^{\dagger}\hat{a} \right).
\end{equation*}
\noindent The next important step was done by N.\,M.~Krylov and N.\,N.~Bogolubov in Ref.~\cite{Bogolubov1939}, see formula (147) which in the case of oscillators takes the form
\begin{align*}
\dot{\rho}_t = -i\omega [\hat{a}^{\dagger}\hat{a}, \rho_t] + &\gamma N \left(\hat{a} \rho_t \hat{a}^{\dagger} - \frac12  \hat{a}^{\dagger}\hat{a} \rho_t - \frac12 \rho_t \hat{a}^{\dagger}\hat{a} \right) \\
+ &\gamma N \left(\hat{a}^{\dagger} \rho_t \hat{a} - \frac12 \hat{a}\hat{a}^{\dagger} \rho_t - \frac12 \rho_t \hat{a}\hat{a}^{\dagger} \right).
\end{align*}
In fact it is a high temperature limit ($ N \gg 1 $) of the widely known model of a damped oscillator (see Ref.~\cite{gardiner2004quantum}, Subsec.~6.1.1 or Ref.~\cite{Bro10}, Subsec.~3.4.6)
\begin{align*}
\dot{\rho}_t = -i\omega [\hat{a}^{\dagger}\hat{a}, \rho_t] +
\gamma (N+1) &\left(\hat{a} \rho_t \hat{a}^{\dagger} - \frac12  \hat{a}^{\dagger}\hat{a} \rho_t - \frac12 \rho_t \hat{a}^{\dagger}\hat{a} \right) \nonumber \\
+ \gamma N &\left(\hat{a}^{\dagger} \rho_t \hat{a} - \frac12 \hat{a}\hat{a}^{\dagger} \rho_t - \frac12 \rho_t \hat{a}\hat{a}^{\dagger} \right).
\end{align*}
At arbitrary temperatures in the explicit form this model was studied by R.~Bausch and A.~Stahl in Ref.~\cite{bausch1967description}.

We consider generators of GKSL form. It was shown in Ref.~\cite{gorini1976completely} for systems in finite-dimensional Hilbert space and in Ref.~\cite{lindblad1976generators} for the case of the bounded generator in the infinite-dimensional Hilbert space that this is the most general form of completely positive trace-preserving semigroup generators. The equation for the density matrix $ \rho_t $ with a GKSL generator has the form
\begin{equation*}
\dot{\rho}_t=-i[\hat{H} , \rho_t] + \sum_i \left( \hat{C}^{(i)} \rho_t (\hat{C}^{(i)})^{\dagger}  - \frac{1}{2} (\hat{C}^{(i)})^{\dagger} \hat{C}^{(i)} \rho_t - \frac{1}{2} \rho_t (\hat{C}^{(i)})^{\dagger} \hat{C}^{(i)} \right),
\end{equation*}
where $ \hat{H} = \hat{H}^{\dagger} $ and $ \hat{C}^{(i)} $ are operators. We consider the case where $ \hat{H} $  are quadratic and $ \hat{C}^{(i)} $ are linear in creation and annihilation operators. GKSL equations have the clearest physical meaning and mathematical properties. They describe quantum Markovian dynamics and could be derived from microscopical equations in some important cases  (see Ref.~\cite{Bro10}, Sec.~3.2). A brief historical review of GKSL equations could be found in Ref.~\cite{chruscinski2017brief}. The mathematically rigorous theory of derivation of GKSL-type equations from the stochastic limit of full Hamiltonian evolutions in the weak coupling and low
density cases was developed by L.~Accardi, A.~Frigerio, Yu.\,G.~Lu, I.\,V.~Volovich at el. (see Refs.~\cite{accardi2002lectures}--\cite{Pech04} for these results and wider bibliography). 
Some important results for the derivation of the weak coupling limit of reduced evolutions by
projective techniques were obtained by E.\,B.~Davies (see Ref.~\cite{davies1976quantum},  Ch.~10)
in the weak coupling limit case and by D\"umcke in the low density case \cite{Dumcke85}. 
Interesting results for derivation of GKSL generators based on non-Wiener quantum stochastic equations were recently obtained by A.\,M.~Basharov \cite{basharov2011spontaneous, Basharov2012, Basharov2014}. The modern view on derivation of GKSL equations and the main directions for future studies in this area could be found in the L.~Accardi review \cite{Accardi17}. GKSL equations also have very natural thermodynamics features \cite{gemmer200419, kosloff2013quantum}. Some important results in this area could be found in works by  I.\,V.~Volovich and A.\,S.~Trushechkin \cite{trushechkin2016perturbative, Tru2017}.

Multimodal bosonic evolution with GKSL-type generators in the Wigner representation was studied by V.\,I.~Man'ko, V.\,V.~Dodonov, O.\,V.~Man'ko \cite{Manko87, DodMan86, dodonov1985quantum}. Some specific examples of such equations were studied in Refs.~\cite{sandulescu1987open}--\cite{isar2013entanglement}. In some special cases multi--modal bosonic GKSL-type equations are solved in Ref.~\cite{Prosen10}. The equilibrium state for coupled oscillators interacting with heat baths at different temperatures were considered by R.\,J.~Glauber and V.\,I.~Man'ko \cite{glauber1984damping}.

One-parameter semigroups correspondent to the quadratic generators are closely related with bosonic and fermionic Gaussian channels which play an important role in quantum information theory.\cite{Holevo15, Parth15, Grepl13} Modern results in this area could be found in Refs.~\cite{holevo15gaussopt}--\cite{Hol17} by A.\,S.~Holevo. It should be mentioned that not every  channel could be represented as a one-parameter semigroup at fixed time.\cite{Cubitt12} In the bosonic case such a possibility was discussed by T.~Heinosaari, A.\,S.~Holevo and M.\,M.~Wolf\cite{heinosaari2009semigroup}.

Modern interest in fermionic Gaussian states originates in Ref.~\cite{Cahill99} by K.\,E.~Cahill and R.\,J.~Glauber. Irreversible fermionic quantum dynamics was considered in Refs.~\cite{Prosen10a}, \cite{Clarc10}. In Ref.~\cite{Prosen10a} it was done by an analog of canonical transformations. In Ref.~\cite{Clarc10} the unitary dilation was used. See further discussion of these results in Ref.~\cite{Teretenkov17b}.

The multi--modal fermionic GKSL equation could also be treated as finite-dimensional matrix equation and solved by general methods for such equations \cite{Alicki07}. However, such an approach needs a solution of systems of linear equations, the number of which depends exponentially on $n$. In this work we obtain the linear equations, the number of which grows linearly in  $n$.

Quadratic fermionic GKSL evolution is closely related to the quantum Gaussian fermionic channels. Ref.~\cite{Bravyi04} by S.~Bravyi,  where the representation of these channels by integral nuclei in fermionic variables was discussed, should be noted here. This work developed the concept of ''fermionic linear optics'' introduced in Refs. \cite{knill2001fermionic}, \cite{terhal2002classical}. Before it the application of quadratic fermionic generators in the quantum information was also studied by A.\,Yu.~Kitaev\cite{kitaev2001unpaired}. The connection between the calculations with fermionic modes and qubit calculations was discussed in Ref.~\cite{bravyi2002fermionic}.  The systematic discussion of fermionic Gaussian channels and literature review could be found in Ref.~\cite{Grepl13}.

\section{Bosonic case}\label{sec:bosonic}

We consider the Hilbert space $\otimes_{j=1}^n\ell_2$. One could define $n$ pairs of creation and annihilation operators acting in such a space (see, for example, Ref.~\cite{scalli2003}, Sec.~1.1.2)
\begin{align*}
\hat{a}_i | \nu_1, \ldots, \nu_i, \ldots, \nu_n \rangle &= \sqrt{\nu_i}| \nu_1, \ldots, \nu_i-1, \ldots, \nu_n \rangle,\\
\hat{a}_i^{\dagger} | \nu_1, \ldots, \nu_i, \ldots, \nu_n \rangle &= \sqrt{\nu_i + 1}| \nu_1, \ldots, \nu_i+1, \ldots, \nu_n \rangle,
\end{align*}
where $ | \nu_1, \ldots, \nu_i, \ldots, \nu_n \rangle, \; \nu_i \in \mathbb{Z}_+$ is a certain basis in $\otimes_{j=1}^n\ell_2$.
Such operators satisfy canonical commutation relations (CCRs)
\begin{equation*}
[\hat{a}_i, \hat{a}_j^{\dagger}] = \delta_{ij} I_{\ell}, \quad [\hat{a}_i, \hat{a}_j] = [\hat{a}_i^{\dagger}, \hat{a}_j^{\dagger}]= 0,
\end{equation*}
where $I_{\ell}$ is an identity matrix in $\otimes_{j=1}^n\ell_2$. For simplicity hereinafter we write just $ \lambda $ instead of $\lambda I_{\ell}$, where $ \lambda \in \mathbb{C} $, for example $ [\hat{a}_i, \hat{a}_j^{\dagger}] = \delta_{ij} $. For  operators from $ \otimes_{j=1}^n\ell_2 $ we denote hermitian conjugation by $ {}^{\dagger} $.

As we work with quadratic and linear combinations of these operators, then it is useful to define the $2n$-dimensional vector $\mathfrak{a} = \left(\hat{a}_1, \cdots, \hat{a}_n, \hat{a}_1^{\dagger}, \cdots, \hat{a}_n^{\dagger} \right)^T$ of annihilation and creation operators and
\begin{align*}
f^T \mathfrak{a} &\equiv \mathfrak{a}^T f \equiv  \sum_{i=1}^n( f_i \hat{a}_i + f_{i+n-1} \hat{a}_i^{\dagger} ), \qquad \forall f \in \mathbb{C}^{2n},\\
\mathfrak{a}^T K \mathfrak{a} &\equiv \sum_{i=1}^n \sum_{j=1}^n( K_{i,j} \hat{a}_i \hat{a}_j + K_{i+n,j} \hat{a}_i^{\dagger} \hat{a}_j + K_{i,j+n} \hat{a}_i \hat{a}_j^{\dagger} + K_{i+n,j+n} \hat{a}_i^{\dagger} \hat{a}_j^{\dagger}), \;  \forall K \in \mathbb{C}^{2n \times 2n}.
\end{align*}
Let us emphasize that this definition does not assume the normal ordering. We denote  the normally ordered form by the colons, e.g.
\begin{equation*}
:\mathfrak{a}^T K \mathfrak{a}: \equiv \sum_{i=1}^n \sum_{j=1}^n( K_{i,j} \hat{a}_i \hat{a}_j + (K_{i+n,j} + K_{i,j+n}) \hat{a}_i^{\dagger} \hat{a}_j  + K_{i+n,j+n} \hat{a}_i^{\dagger} \hat{a}_j^{\dagger}), \qquad \forall K \in \mathbb{C}^{2n \times 2n}.
\end{equation*}
The hermitian conjugation of the matrices $  K \in \mathbb{C}^{2n \times 2n} $ we denote by $ {}^+ $.  Let us also introduce the following $2n \times 2n$-matrices:
\begin{equation}\label{JEIdef}
J = \left(
\begin{array}{cc}
0 & -I_n \\
I_n & 0
\end{array}
\right), \qquad
E = \left(
\begin{array}{cc}
0 & I_n \\
I_n & 0
\end{array}
\right),
\qquad
I = \left(
\begin{array}{cc}
I_n & 0  \\
0 & I_n
\end{array}
\right),
\end{equation}
where $ I_n $ is an identity matrix in $ \mathbb{C}^{n \times n} $. Then the CCRs take the form
\begin{equation*}
[f^T \mathfrak{a},\mathfrak{a}^T g]  = - f^T J g, \quad \forall g, f \in \mathbb{C}^{2n}.
\end{equation*}

Let us introduce the following definition in order to express the hermitian conjugation of the quadratic and linear forms in terms of their coefficients.
\begin{definition}
	\label{tildeDef}
	$\sim$-conjugation of vectors and matrices is defined by the following formulae
	\begin{equation*}
	\tilde{g} = E\overline{g}, \; g \in \mathbb{C}^{2n}, \qquad \tilde{K} = E \overline{K} E, \; K \in \mathbb{C}^{2n \times 2n},
	\end{equation*}
	where the overline is an (elementwise) complex conjugation and the matrix $E$ is defined by \eqref{JEIdef}.
\end{definition}

Then the hermitian conjugation of the quadratic and linear forms could be calculated by the formulae $ (g^T \mathfrak{a})^{\dagger} =  \tilde{g}^T \mathfrak{a}$,  $(\mathfrak{a}^T K \mathfrak{a})^{\dagger}=\mathfrak{a}^T\tilde{K}^T \mathfrak{a}$,  $\forall g \in \mathbb{C}^{2n}, K \in \mathbb{C}^{2n \times 2n} $. In particular, $(\mathfrak{a}^T K \mathfrak{a})^{\dagger}=\mathfrak{a}^T\tilde{K} \mathfrak{a}$ in the case when $K = K^T$. Let us note that $\sim$-conjugation is an involution  $ \tilde{\tilde{K}} = K, \tilde{\tilde{g}} = g, K \in \mathbb{C}^{2n \times 2n} $
and it is semi-linear i.e.
\begin{equation*}
(\lambda_1 K_1 + \lambda_2 K_2)^{\sim} = \overline{\lambda}_1 \tilde{K}_1 + \overline{\lambda}_2 \tilde{K}_2, \qquad \forall K_1, K_2 \in \mathbb{C}^{2n \times 2n}, \quad \lambda_1, \lambda_2 \in \mathbb{C}.
\end{equation*}
In addition, let us mention that  $\tilde{J} = - J$ and $\tilde{E} = E$.

The set of the trace-class operators in $\otimes_{j=1}^n\ell_2$ we denote by $\mathcal{T}(\otimes_{j=1}^n\ell_2)$. We call a self-adjoint non-negatively defined operator with trace $ 1 $ \textit{a density matrix}.  We use the notation $ \mathrm{tr} \; $ for the trace with respect to $\otimes_{j=1}^n\ell_2$ and $ \mathrm{Tr} \; $ for the one with respect to $ \mathbb{C}^{2n} $.

\subsection{Basic calculations with quadratic and linear forms}
\label{subsec:basicBos}

In this subsection we present basic formula for computations with quadratic and linear forms in bosonic creation and annihilation operators. It could be used as reference information even out of the context of our article. The proofs of them could be found in Refs.~\cite{Teretenkov16} and \cite{Teretenkov17}.

\begin{lemma}\label{lemmaS}(Symmetrization)
	Let $K \in \mathbb{C}^{2n \times 2n}$, then
	\begin{equation*}
	\mathfrak{a}^T K\mathfrak{a} = \mathfrak{a}^T \frac12(K+K^T) \mathfrak{a} - \frac12 \mathrm{Tr} \; K^T J = \mathfrak{a}^T \frac12(K+K^T) \mathfrak{a} + \frac12 \mathrm{Tr} \; K J.
	\end{equation*}
\end{lemma}

\begin{lemma}\label{lemmaN} (Normal ordering)
	Let $K= K^T \in \mathbb{C}^{2n \times 2n}$, then
	\begin{equation*}
	\mathfrak{a}^T K \mathfrak{a} = :\mathfrak{a}^T K \mathfrak{a} : + \frac12 \mathrm{Tr} \; E K.
	\end{equation*}
\end{lemma}

\begin{lemma}\label{lemma1}(Commutation relations for linear and quadratic forms)
	Let $K=K^T, M=M^T \in \mathbb{C}^{2n \times 2n}$ and $g, f \in \mathbb{C}^{2n} $, then:
	\begin{equation*}
	\left[\frac12 \mathfrak{a}^T K \mathfrak{a} + g^T\mathfrak{a}, \frac12 \mathfrak{a}^T M \mathfrak{a} + f^T\mathfrak{a}\right]= \frac12 \mathfrak{a}^T( MJK - KJM) \mathfrak{a}+(f^TJK - g^TJM)\mathfrak{a} - g^T J f.
	\end{equation*}
\end{lemma}

\begin{lemma}\label{lemma2}(Quadratic and linear forms sandwiched by exponentials.)
	Let $K=K^T, M=M^T \in \mathbb{C}^{2n \times 2n}$ and $g, f \in \mathbb{C}^{2n} $, then:
	\begin{align*}
	&e^{\frac12 \mathfrak{a}^T K \mathfrak{a} + g^T\mathfrak{a} } \left(\frac12 \mathfrak{a}^T M \mathfrak{a} + f^T\mathfrak{a} \right) e^{- \frac12 \mathfrak{a}^T K \mathfrak{a} - g^T\mathfrak{a}} =
	\\
	&\;= \frac12 \mathfrak{a}^T e^{-K J } M e^{J K} \mathfrak{a}  + \left( e^{-K J } M \frac{e^{J K} - I}{J K} J g  + e^{-K J }  f\right)^T  \mathfrak{a} \;+
	\\
	&\quad+ \left( \frac12 g^T J \frac{e^{-K J} - I}{KJ} M  + f^T\right)\frac{e^{J K} - I}{J K} J g.
	\end{align*}
\end{lemma}

\begin{lemma}\label{lemma3}
	(Derivative of exponential.)
	Let $K_t = K_t^T \in \mathbb{C}^{2n \times 2n}$ be differentiable function with respect to $ t $ for $ t \in \mathbb{R}_+ $, then:
	\begin{align*}
	\left( \frac{d}{dt}e^{\frac12 \mathfrak{a}^T K_t \mathfrak{a} + g_t^T\mathfrak{a} + c_t} \right) e^{- \frac12 \mathfrak{a}^T K_t \mathfrak{a} -g_t^T\mathfrak{a} - c_t} =
	\frac12 \mathfrak{a}^T  e^{-K_t J} \frac{d}{dt} e^{K_t J} J^{-1} \mathfrak{a} \\+ \mathfrak{a}^T  e^{-K_t J} \frac{d}{dt} \left( \frac{e^{K_t J } - I}{K_tJ}  g_t \right) + \frac12 g_t^T J \frac{e^{-K_t J} - I}{K_t J} \frac{d}{dt} \left( \frac{e^{K_t J} - I}{K_t J} g_t \right) \\+
	\frac{d}{dt} \left(\frac12  g_t^T J \frac{1}{K_t J} \left( \sinh(K_t J)  - K_t J\right) \frac{1}{K_t J} g_t + c_t\right)
	\end{align*}
	and the matrix $e^{-K_t J} \frac{d}{dt} e^{K_t J} J^{-1}$ is symmetric.
\end{lemma}
This result is based on the Feynman-Wilcox formula\cite{Chebotarev12}.

\subsection{Quadratic superoperators}
\label{subsec:QuadBos}

In this subsection we define the GKSL generators which are quadratic in bosonic creation and annihilation operators.

\begin{definition}
	We call an unbounded operator $\mathcal{L}: \mathcal{T}(\otimes_{j=1}^n\ell_2) \rightarrow \mathcal{T}(\otimes_{j=1}^n\ell_2) $ of the form
	\begin{equation}\label{quad_gen}
	\mathcal{L}(\hat{X}) = \mathfrak{a}^T \Gamma \hat{X}  \mathfrak{a} + \mathfrak{a}^T \Gamma_L  \mathfrak{a} \hat{X}  + \hat{X}  \mathfrak{a}^T \Gamma_R  \mathfrak{a} + \hat{X}  g_R^T \mathfrak{a} + g_L^T \mathfrak{a} \hat{X}  + \lambda \hat{X} ,
	\end{equation}
	where $\Gamma, \Gamma_L = \Gamma_L^T, \Gamma_R = \Gamma_R^T\in \mathbb{C}^{2n \times 2n}, g_L, g_R \in \mathbb{C}^{2n}, \lambda \in \mathbb{C}$, \textit{a quadratic bosonic superoperator}.
\end{definition}

\begin{definition}
	Let  $\mathcal{L}$ be a quadratic superoperator of form \eqref{quad_gen}. Let us define a superoperator
	$\mathcal{L}^*$ by the formula
	\begin{equation*}
	\mathrm{tr} \; ((\mathcal{L}^*(\hat{X}))^{\dagger} \rho ) = \mathrm{tr} \; (\hat{X}^{\dagger} \mathcal{L}(\rho)),
	\end{equation*}
	where $ \rho $ is a density matrix and $ \mathrm{tr} \; ( \mathfrak{a} \rho )< \infty  $, $ \mathrm{tr} \;  (\mathfrak{a} \mathfrak{a}^T \rho) < \infty  $, $ \hat{X} $ is a (possibly unbounded) operator in $ \otimes_{j=1}^n\ell_2 $.
\end{definition}

\begin{definition}
	Let $ \mathcal{L}: \mathcal{T}(\otimes_{j=1}^n\ell_2) \rightarrow \mathcal{T}(\otimes_{j=1}^n\ell_2) $ be a quadratic superoperator of form \eqref{quad_gen}. Let us define a superoperator by the formula
	\begin{equation*}
	\mathcal{L}^+(\rho) = \left( \mathcal{L}(\rho^{\dagger}) \right)^{\dagger},
	\end{equation*}
	
\end{definition}

\begin{lemma}\label{plus_conj}
	Let $\mathcal{L}$ be a quadratic superoperator of form \eqref{quad_gen}, then
	\begin{align*}
	\mathcal{L}^+(\rho) &= \mathfrak{a}^T \tilde{\Gamma}^T \rho \mathfrak{a} + \mathfrak{a}^T \tilde{\Gamma}_R  \mathfrak{a} \rho + \rho \mathfrak{a}^T \tilde{\Gamma}_L \mathfrak{a} + \rho \tilde{g}_L^T \mathfrak{a} + \tilde{g}_R^T \mathfrak{a} \rho +\overline{ \lambda} \rho,
	\\
	\mathcal{L}^*(\hat{X}) &= \mathfrak{a}^T \tilde{\Gamma} \hat{X} \mathfrak{a} + \mathfrak{a}^T \tilde{\Gamma}_L  \mathfrak{a} \hat{X} + \hat{X} \mathfrak{a}^T \tilde{\Gamma}_R \mathfrak{a} + \hat{X} \tilde{g}_R^T \mathfrak{a} + \tilde{g}_L^T \mathfrak{a} \hat{X} +\overline{ \lambda} \hat{X}.
	\end{align*}
\end{lemma}

Let us note that the coefficients $\Gamma, \Gamma_L = \Gamma_L^T, \Gamma_R = \Gamma_R^T\in \mathbb{C}^{2n \times 2n}, g_L, g_R \in \mathbb{C}^{2n}, \lambda \in \mathbb{C}$ of superoperator \eqref{quad_gen} are uniquely defined.

\begin{lemma}
	The quadratic superoperator $\mathcal{L}$ of form \eqref{quad_gen} satisfies the  conditions $\mathcal{L}^+ = \mathcal{L}$ and  $\mathcal{L}^*(I) = 0$  on its domain if and only if
	\begin{equation}\label{pos_gen}
	\mathcal{L}(\rho) = \mathcal{L}_{H, \Gamma, f}(\rho) \equiv -i\left[\frac{1}{2} \mathfrak{a}^T H \mathfrak{a} + f^T \mathfrak{a} , \rho \right] +  \mathfrak{a}^T \rho \Gamma \mathfrak{a}  - \frac{1}{2} \mathfrak{a}^T \Gamma^T \mathfrak{a} \rho  - \frac{1}{2} \rho \; \mathfrak{a}^T \Gamma^T \mathfrak{a} ,
	\end{equation}
	where $ H = H^T = \tilde{H} $, $ \Gamma^T = \tilde{\Gamma} $, $ f = \tilde{f} $.
\end{lemma}

We are interested in GKSL generators of the form
\begin{equation}\label{lind_gen1}
\mathcal{L}(\rho)=-i[\hat{H} , \rho] + \sum_i \left( \hat{C}^{(i)} \rho (\hat{C}^{(i)})^{\dagger}  - \frac{1}{2} (\hat{C}^{(i)})^{\dagger} \hat{C}^{(i)} \rho - \frac{1}{2} \rho (\hat{C}^{(i)})^{\dagger} \hat{C}^{(i)} \right),
\end{equation}
where $\hat{H}$ contains quadratic and linear terms in creation and annihilation operators, $\hat{C}_i$ contains linear and scalar terms. Let us stress that actually both theorems from Refs.~\cite{gorini1976completely} and \cite{lindblad1976generators} are not applicable to the case of unbounded generators, moreover, actually interesting non-GKSL examples are presented in literature.\cite{Holevo18}
\begin{lemma}\label{lemmaLindbladform}
	Let $\hat{H} = \frac12 \mathfrak{a}^T H \mathfrak{a} + h^T \mathfrak{a}$ ($ H = H^T = \tilde{H}$, $ h = \tilde{h} $) and $\hat{C}^{(i)}= \gamma_i^T \mathfrak{a} + c_i$, then generator \eqref{lind_gen1} takes the form \eqref{pos_gen}
	where
	\begin{equation}\label{lind_param}
	\Gamma = \sum\limits_i \gamma_i \tilde{\gamma}_i^T, \qquad l = \sum\limits_i \overline{c_i} \gamma_i, \qquad f = h + i \; \frac{l - \tilde{l}}{2}.
	\end{equation}
	Moreover, $\Gamma$ and $f$  can be presented in form \eqref{lind_param} if and only if
	\begin{equation}\label{lind_param1}
	\Gamma^T = \tilde{\Gamma}, \quad \Gamma E \geqslant 0, \quad f = \tilde{f}.
	\end{equation}
\end{lemma}

Thus, all linear terms could be absorbed by $\hat{H}$ and $c_i$ could be omitted. Actually, it could be done for arbitrary GKSL generators (see Ref.~\cite{Cheb2000}, p.~70 or Ref.~\cite{Bro10}, p.~123).

THe quadratic GKSL generator is a special case of  \eqref{pos_gen} with the additional condition $\Gamma E \geqslant 0$.  But in general case \eqref{pos_gen} the condition $(\Gamma E)^+ = \Gamma E $ allows one to decompose $\Gamma E = \sum\limits_i \gamma_i \gamma_i^+ - \sum\limits_j \beta_j \beta_j^+$ and, hence, to represent \eqref{pos_gen} as a difference of two GKSL generators of form \eqref{pos_gen}--\eqref{lind_param1} with $\Gamma_1 = \sum_i \gamma_i \tilde{\gamma}_i^T$ and  $\Gamma_2 = \sum_j \beta_j \tilde{\beta}_j^T$. The terms of  $\hat{H}$ could be absorbed by the first as well as by the second generator.

The generators \eqref{lind_gen1} can coincide for different $\hat{C}^{(i)}$, but form \eqref{pos_gen} is unique. The fact that an arbitrary non-negative definite Hermitian matrix could be diagonalized allows one to decompose $\Gamma E = \sum\limits_{i=1}^{2n} \gamma_i \gamma_i^+$. So one could represent the GKSL generator in form \eqref{lind_gen1}, where $i$ takes no more than $2 n$ values.

In the case $\hat{C}^{(i)} = (\hat{C}^{(i)})^{\dagger}$ the GKSL generator takes the form
\begin{equation*}
\mathcal{L}(\rho)=-i[\hat{H} , \rho] - \frac12 \sum_i [\hat{C}^{(i)}, [\hat{C}^{(i)}, \rho]].
\end{equation*}
Such generators are closely related to the classical diffusion. First of all this is a direct analog of the generator of the classical Fokker--Planck equation, where the adjacent action of $\hat{H}, \hat{C}^{(i)} $, i.e.  $ [\hat{H}, \cdot] $ and $ [\hat{C}^{(i)}, \cdot] $, plays the role of the derivatives.  Moreover, the terms  $ \frac12 [\hat{C}^{(i)}, [\hat{C}^{(i)}, \rho]] $ could arise as averaging with respect to the (classical) Wiener process (see Ref.~\cite{Holevo2003}, p.~82). The matrix $\Gamma$ in \eqref{lind_param1} satisfies the condition $\Gamma = \Gamma^T$ in this case.

\begin{lemma}
	\label{lemma:adjointSuperOp}
	The adjoint superoperator for \eqref{pos_gen} has the form
	\begin{equation*}
	\mathcal{L}_{H, \Gamma, f}^* (\hat{X}) = i\left[\frac{1}{2} \mathfrak{a}^T H \mathfrak{a} + f^T \mathfrak{a} , \hat{X} \right] +  \mathfrak{a}^T \hat{X} \Gamma^T \mathfrak{a}  - \frac{1}{2} \mathfrak{a}^T \Gamma^T \mathfrak{a} \hat{X}  - \frac{1}{2} \hat{X} \; \mathfrak{a}^T \Gamma^T \mathfrak{a}, \hat{X} \in \mathcal{B}(\otimes_{j=1}^n\ell_2).
	\end{equation*}
\end{lemma}

The adjoint GKSL quadratic bosonic generators themselves are GKSL quadratic bosonic generators up to a constant.

\begin{corollary}
	Let $ \rho  $ be a density matrix, then the adjoint generator $ \mathcal{L}_{H, \Gamma, f}^* $ is related to $ \mathcal{L}_{-H, \Gamma^T, -f}  $ by the formula
	\begin{equation*}
	\mathcal{L}_{H, \Gamma, f}^* (\rho) = \mathcal{L}_{-H, \Gamma^T, -f} (\rho) - \frac12 (\mathrm{Tr} \; \Gamma^T J) \rho.
	\end{equation*}
\end{corollary}

The following proposition is not used further but it could be used to investigate the dissipative analogs of conservation laws.

\begin{proposition}
	\label{prop:commut}
	The commutator of two quadratic superoperators of form \eqref{pos_gen} on appropriate domain could be defined by the formula
	\begin{equation*}
	[\mathcal{L}_{H_1, \Gamma_1, f_1},
	\mathcal{L}_{H_2, \Gamma_2, f_2}] = \mathcal{L}_{H_{12}, \Gamma_{12}, f_{12}},
	\end{equation*}
	where
	\begin{align*}
	H_{12} =& i \left( [H_1 J, H_2 J]  +  \left[\frac{\Gamma_2 - \Gamma_2^T}{2} J, \frac{\Gamma_1 - \Gamma_1^T}{2} J\right] \right) J,\\
	\Gamma_{12} =& i ([H_1 J, \Gamma_2 J] - [H_2 J, \Gamma_1 J]  + \frac{i}{2} (\{\Gamma_1^T J, \Gamma_2 J\} -\{ \Gamma_1 J, \Gamma_2^T J \})) J,\\
	f_{12} =&  i \left( H_1 + i \frac{\Gamma_1-\Gamma_1^T}{2} \right) J f_2 - \left( H_2 +i \frac{\Gamma_2-\Gamma_2^T}{2} \right) J f_1  .
	\end{align*}
\end{proposition}
Thus, superoperators \eqref{pos_gen} form a Lie algebra. For further study it is interesting to understand relation between exact solutions of GKSL equations and Lie algebras of superoperators of certain form.

\subsection{Gaussian states}
\label{subsec:GaussBos}

In this section we define the Gaussian states and represent them in different forms which are usually used in literature.

\begin{definition}
	Let us define the \textit{mean vector} $m$ and \textit{the matrix  $D$ of the second central moments} for the density matrix  by the formulae
	\begin{equation}
	\label{meanSecMomDef}
	m = \mathrm{tr} \; (\mathfrak{a} \, \rho), \qquad D =  \mathrm{tr} \; \left( (\mathfrak{a} - m) \; (\mathfrak{a} -m)^T \rho \right).
	\end{equation}
\end{definition}

\begin{definition}
	\label{covMatDef0}
	We call the symmetric part of the matrix of the second central moments
	\begin{equation}\label{covMatDef}
	C = \frac{D + D^T}{2}.
	\end{equation}
	\textit{the covariance matrix}.
\end{definition}

The antisymmetric part of the matrix $ D $ is defined by CCRs as $ D - D^T = - J $ and carries no information about the state. The matrix  $ D $ could be uniquely defined by $  C $ as $ D = C - J/2 $.

\begin{definition}
	We call the function
	\begin{equation*}
	\chi_W (\mathbf{z}) \equiv \mathrm{tr} \; (\rho \; e^{i \mathbf{z}^T \mathfrak{a}})
	\end{equation*}
	the \textit{characteristic function}.
\end{definition}

This function is defined for every density matrix $ \rho $ and uniquely defines this matrix  (see Ref.~\cite{Holevo15}, Subsec.~12.3.1).

\begin{definition}
	A density matrix $ \rho $ is  called \textit{Gaussian state} if its characteristic function has the form
	\begin{equation}\label{eq:charW}
	\chi_W (\mathbf{z})  = e^{- \frac12 \mathbf{z}^T C \mathbf{z} + i   \mathbf{z}^T m}.
	\end{equation}
\end{definition}

One could also define the Gaussian state as a state the characteristic function $ \chi_W (\mathbf{z}) $ of which is exponential of sum of linear and quadratic forms. And only then one could prove that its coefficients are exactly the mean vector and the covariance matrix defined by \eqref{meanSecMomDef} and \eqref{covMatDef}. But we have already absorbed this fact in our definition. Thus, any Gaussian state is uniquely defined by $ m $ and $ C $ (or by $ m $ and $ D $). But it is natural to ask what conditions are imposed on $ m $ and $ D $.

\begin{proposition}\label{propSecMomMatCond}
	The vector $ m \in \mathbb{C}^{2n} $ and the matrix $ D \in \mathbb{C}^{2n \times 2n} $ satisfy the conditions
	\begin{equation}\label{commD}
	m = \tilde{m}, \qquad D^T = \tilde{D}, \qquad D E \geqslant 0, D - D^T = - J
	\end{equation}
	if and only if there exists a (unique) Gaussian state with the mean vector $m$ and the matrix of the second central moments $D$.
\end{proposition}

The proof of this proposition could be found on p.~286 in Ref.~\cite{Holevo15}. Moreover, conditions \eqref{commD} are held for arbitrary states for which moments \eqref{meanSecMomDef} are defined.

The subscript $ W $ of the function $ \chi_W (\mathbf{z}) $ refers to the Weyl-ordering. Normally and anti-normally ordered characteristic functions are also frequently used (see Ref.~\cite{carmichael2000statistical}, Ch.~4). In the case of Gaussian states they take the form
\begin{align*}
\chi_N (\mathbf{z}) &\equiv \mathrm{tr} \; (\rho \; e^{i \mathbf{z}^T \frac{I + EJ}{2} \mathfrak{a}} e^{i \mathbf{z}^T \frac{I - EJ}{2} \mathfrak{a}}) = e^{- \frac12 \mathbf{z}^T \left( C + \frac12 E \right)  \mathbf{z} + i   \mathbf{z}^T m},
\\
\chi_A (\mathbf{z}) &\equiv \mathrm{tr} \; (\rho \; e^{i \mathbf{z}^T \frac{I - EJ}{2} \mathfrak{a}} e^{i \mathbf{z}^T \frac{I + EJ}{2} \mathfrak{a}}) =  e^{- \frac12 \mathbf{z}^T \left( C - \frac12 E \right)  \mathbf{z} + i   \mathbf{z}^T m}.
\end{align*}

Non-pure Gaussain states could be also represented in the exponential form
\begin{equation}\label{state}
\rho = e^{\frac12 \mathfrak{a}^T K \mathfrak{a} + g^T\mathfrak{a} + c},
\end{equation}
where $K=K^T = \tilde{K} \in \mathbb{C}^{2n \times 2n}$ is such that $K E < 0$, $g = \tilde{g} \in \mathbb{C}^{2n}$ and
\begin{equation}\label{c_from_norm}
e^{s} = \sqrt{ |\det \left( e^{K J} - I \right)|} \;e^{\frac12 g^T \frac{1}{K} g}
\end{equation}
in accordance with results on p. 289 in Ref.~\cite{Holevo15} in our notation.
Formula \eqref{c_from_norm} is mainly based on the following result (see Ref.~\cite{Holevo15}, p.~280 or Ref.~\cite{colpa1978diagonalization}).
\begin{proposition}
	\label{prop:bidiag}
	Let $K=K^T = \tilde{K} \in \mathbb{C}^{2n \times 2n}$ and $ K E < 0 $, then there exists a~symplectic matrix $ S $ ($ S J S^T = J$, $ S = \tilde{S}  $) such that $  K $ takes the bidiagonal form
	\begin{equation*}
	K = - S
	\begin{pmatrix}
	0 & \Lambda\\
	\Lambda & 0
	\end{pmatrix}
	S^T,
	\end{equation*}
	where $ \Lambda = \mathrm{diag}\; (\lambda_1, \ldots, \lambda_n)$ is a real diagonal matrix with strictly positive $ \lambda_i >0 $.
\end{proposition}

Form \eqref{state} is related to form \eqref{eq:charW} by the following proposition. It is also the direct result of proposition \ref{prop:bidiag}.
\begin{proposition}
	\label{coroll:secMomMatrixCondNonDeg}
	Let  $K=K^T = \tilde{K} \in \mathbb{C}^{2n \times 2n}$ and $ K E < 0 $ and $ g = \tilde{g} $, then there exists the Gaussian state of form \eqref{state} with the mean vector $m$ and the matrix of the second central moments $D$ defined by the formulae
	\begin{equation*}
	m = - K^{-1} g, \qquad D = J e^{K J} (I - e^{K J})^{-1}.
	\end{equation*}
\end{proposition}

The normally ordered form is also frequently used. It could be obtained by the following proposition. Its proof could be found in Ref.~\cite{Teretenkov17}.

\begin{proposition}\label{theorem2}
	Let $K=K^T = \tilde{K} \in \mathbb{C}^{2n \times 2n}$  be such that $K E < 0$ , $g = \tilde{g} \in \mathbb{C}^{2n}$, and $s \in \mathbb{R}$ is defined by the normalization condition, i.e. by formula \eqref{c_from_norm}, then the normal symbol of the density matrix equals $ e^{\frac12 \mathbf{z}^T R \mathbf{z} + q^T \mathbf{z} + r} $, where
	\begin{equation*}
	R  = - 2 \left(E + J \frac{I + e^{K J}}{I - e^{K J}}\right)^{-1}, \quad q = R K^{-1} g, \quad  e^r = \sqrt{\det (E R)} e^{ \frac12 q^T R^{-1} q}.
	\end{equation*}
\end{proposition}

\subsection{Dynamics of moments and Gaussian solutions}
\label{subsec:DynBos}

In this section we obtain the Gaussian solutions of the equation
\begin{equation}\label{eq:CauchyProbTime}
\frac{d}{dt} \rho_t = \mathcal{L}_{H_t, \Gamma_t, f_t}(\rho_t),
\end{equation}
where $ H_t, \Gamma_t, f_t $ are continuous functions in $ t $ on $ \mathbb{R}_+ $ satisfying the conditions $ H_t = H_t^T = \tilde{H}_t $,	$ \Gamma_t^T = \tilde{\Gamma}_t $,  $ \Gamma_t E \geqslant 0 $,  $ f_t = \tilde{f}_t $ for all $ t \in \mathbb{R}_+ $.

\begin{lemma}
	\label{conjGenAction}
	The action of the adjoint generator in creation and annihilation operators and their products is defined by the formulae
	\begin{align*}
	\mathcal{L}_{H, \Gamma, f}^* \left( \mathfrak{a} \right) &=  J \left(i H + \frac{\Gamma^T - \Gamma}{2}\right) \mathfrak{a} + i J f,\\
	\mathcal{L}_{H, \Gamma, f}^* \left( \mathfrak{a} \; \mathfrak{a}^T \right) &=  J \left(i H + \frac{\Gamma^T - \Gamma}{2} \right) \mathfrak{a} \; \mathfrak{a}^T + \mathfrak{a} \; \mathfrak{a}^T \left(-i H + \frac{\Gamma^T - \Gamma}{2} \right) J + J \Gamma^T J  \\
	& \qquad \qquad\qquad\qquad\qquad\quad\; + i J f \; \mathfrak{a}^T - i \mathfrak{a} \; f^T J.
	\end{align*}
\end{lemma}

This lemma is a direct corollary of lemma \ref{lemma:adjointSuperOp}.

\begin{theorem}
	\label{firstSecMomCor}
	Let the density matrix $ \rho_t $ satisfy Eq.~\eqref{eq:CauchyProbTime} and have the finite moments $ m_t $ and the second central moments $ D_t $, then the mean vector and the second central moments satisfy
	\begin{align}\label{eq:firstMoment}
	\frac{d}{dt} m_t &= J \left(i H_t + \frac{\Gamma_t^T - \Gamma_t}{2}\right) m_t  + i J f_t,
	\\
	\label{eq:secondMoments}
	\frac{d}{dt} D_t &= J \left(i H_t + \frac{\Gamma_t^T - \Gamma_t}{2}\right) D_t + D_t \left(- i H_t + \frac{\Gamma_t^T - \Gamma_t}{2}\right) J + J \Gamma_t^T J.
	\end{align}
	Here $ m_t $ and $ D_t $ are defined by formulae \eqref{meanSecMomDef}.
\end{theorem}

The proof could be obtained by substituting \eqref{state} into \eqref{eq:CauchyProbTime} and using lemmas~\ref{lemma1}--\ref{lemma3}\cite{Teretenkov16, Teretenkov17} or by averaging formulae from lemma \ref{conjGenAction}.\cite{Teretenkov19} The former approach is applicable only in the case of the Gaussian solutions, while the latter is applicable in the general case of  density matrices with finite first and second moments.

If one symmetrizes Eq.~\eqref{eq:secondMoments}, one obtains the equation for the covariance matrix.

\begin{proposition}
	\label{covMatProp}
	Let the matrix $ D_t $ satisfy Eq.~\eqref{eq:secondMoments},  then the covariance matrix $C_t =(D_t + D_t^T)/2$ satisfies the equation
	\begin{equation}\label{eq:covMat}
	\frac{d}{dt} C_t = J \left(i H_t + \frac{\Gamma_t^T - \Gamma_t}{2}\right) C_t + C_t \left(- i H_t + \frac{\Gamma_t^T - \Gamma_t}{2}\right) J + J \frac{\Gamma_t +\Gamma_t^T}{2} J.
	\end{equation}
	If, in addition,  the matrix $ D_0 $ satisfies the condition $D_0 - D_0^T = - J$, then $D_t - D_t^T = - J$ for all $ t>0 $.
\end{proposition}

The solution of Eqs.~\eqref{eq:firstMoment}, \eqref{eq:secondMoments}, \eqref{eq:covMat} could be expressed in terms of the solution of Eq.~\eqref{eq:firstMoment} without inhomogeneity by the following proposition.

\begin{proposition}
	\label{prop:2nSol}
	Let $ G_t $ be a solution of the Cauchy problem
	\begin{equation*}
	\dot{G}_t = J \left(i H_t + \frac{\Gamma_t^T - \Gamma_t}{2}\right) G_t, \qquad G_0 = I.
	\end{equation*}
	Then the solutions of Eqs.~\eqref{eq:firstMoment}, \eqref{eq:secondMoments}, \eqref{eq:covMat} could be represented in the form
	\begin{align*}
	m_t &= G_t m_0 + i \int_0^t G_{t} G_{t'}^{-1}  J f_{t'} dt',\\
	D_t &= G_t D_0 G_t^T + \int_0^t G_{t} G_{t'}^{-1} J \Gamma_{t'}^T J (G_{t} G_{t'}^{-1})^T dt',\\
	C_t &= G_t C_0 G_t^T +  \int_0^t G_{t} G_{t'}^{-1} J \frac{\Gamma_{t'} + \Gamma_{t'}^T}{2} J (G_{t} G_{t'}^{-1})^T dt'.
	\end{align*}
\end{proposition}

Eqs.~\eqref{eq:secondMoments}, \eqref{eq:covMat} could be considered as linear differential equations in $ (2n)^2 \times (2n)^2 $ variables. The following proposition allows one to represent their solution in terms of solutions of a linear differential equation in $ 4n $ variables. The proof of the  proposition analogous to the following one and further discussion could be found in~Ref.~\cite{Teretenkov17}.

\begin{proposition}\label{doubleEqProp}
	If $ m_t $ satisfies Eq.~\eqref{eq:firstMoment}, then the following differential equation is held
	\begin{equation*}
	\frac{d}{dt}
	\begin{pmatrix}
	m_t \\ 1
	\end{pmatrix}
	=
	\begin{pmatrix}
	J \left(i H_t + \frac{\Gamma_t^T - \Gamma_t}{2}\right) & i J f_t\\
	0^T & 0
	\end{pmatrix}
	\begin{pmatrix}
	m_t \\ 1
	\end{pmatrix}.
	\end{equation*}
	
	The solution of Eq.~\eqref{eq:secondMoments} could be represented in the form  $D_t = X_t Y_t^{-1}$, where the matrices $ X_t $ and $ Y_t $ are defined by the following differential equation
	\begin{equation*}
	\frac{d}{dt}
	\left( \begin{array}{c}
	X_t\\
	Y_t
	\end{array} \right) =
	\left( \begin{array}{cc}
	J \left(i H_t + \frac{\Gamma_t^T - \Gamma_t}{2}\right) & J \Gamma_t^T J\\
	0 &  \left(i H_t - \frac{\Gamma_t^T - \Gamma_t}{2}\right)J
	\end{array} \right)
	\left( \begin{array}{c}
	X_t\\
	Y_t
	\end{array} \right)
	\end{equation*}
	with an arbitrary initial condition such that $ X_0   Y_0^{-1} =  D_0$. In particular, one could assume $ X_0 = D_0, Y_0 = I $.
	
	Similarly, the solution of Eq.~\eqref{eq:covMat} could be represented in the form $C_t = X_t' Y_t^{-1}$,where the matrices $ X_t' $ and $ Y_t $ are defined by the following differential equation
	\begin{equation*}
	\frac{d}{dt}
	\left( \begin{array}{c}
	X_t'\\
	Y_t
	\end{array} \right) =
	\left( \begin{array}{cc}
	J \left(i H_t + \frac{\Gamma_t^T - \Gamma_t}{2}\right) & J \frac{\Gamma_t^T + \Gamma_t}{2} J\\
	0 &  \left(i H_t - \frac{\Gamma_t^T - \Gamma_t}{2}\right)J
	\end{array} \right)
	\left( \begin{array}{c}
	X_t'\\
	Y_t
	\end{array} \right)
	\end{equation*}
	with an arbitrary initial condition such that $ X_0'  Y_0^{-1} =  C_0$. In particular, one could assume $ X_0' = C_0, Y_0 = I $.
\end{proposition}

For the case of constant coefficients proposition \ref{prop:2nSol} could be represented in more explicit form.

\begin{proposition}
	\label{propConsSol}
	If the functions $ H_t, \Gamma_t , f_t$ in \eqref{eq:firstMoment} are constants $ H, \Gamma , f$ and the matrix $ i H + \frac{\Gamma^T - \Gamma}{2} $ is non-degenerate, then the solution of Eq.~\eqref{eq:firstMoment} takes the form
	\begin{equation*}
	m_t = e^{J \left(i H + \frac{\Gamma^T - \Gamma}{2}\right) t} m_0 + i \frac{ e^{J \left(i H + \frac{\Gamma^T - \Gamma}{2}\right) t} - I}{J \left(i H + \frac{\Gamma^T - \Gamma}{2}\right)} J f.
	\end{equation*}	
	If the functions $ H_t, \Gamma_t $ in \eqref{eq:secondMoments} are constants $ H, \Gamma $, then the solution of Eq.~\eqref{eq:secondMoments} takes the form
	\begin{equation*}
	D_t = e^{ J \left(i H + \frac{\Gamma^T - \Gamma}{2}\right) t} D_0 e^{\left(-i H + \frac{\Gamma^T - \Gamma}{2}\right) J t} +   \int_0^t e^{ J \left(i H + \frac{\Gamma^T - \Gamma}{2}\right) t'}  J \Gamma^T J  e^{\left(-i H + \frac{\Gamma^T - \Gamma}{2}\right) J t'} d t'.
	\end{equation*}
	If the functions $ H_t, \Gamma_t $ in \eqref{eq:covMat} are constants $ H, \Gamma $, then the solution of Eq.~\eqref{eq:covMat} takes the form
	\begin{equation*}
	C_t = e^{ J \left(i H + \frac{\Gamma^T - \Gamma}{2}\right) t} C_0 e^{\left(-i H + \frac{\Gamma^T - \Gamma}{2}\right) J t} +   \int_0^t e^{ J \left(i H + \frac{\Gamma^T - \Gamma}{2}\right) t'}  J \frac{\Gamma + \Gamma^T}{2}J  e^{\left(-i H + \frac{\Gamma^T - \Gamma}{2}\right) J t'} d t'.
	\end{equation*}
\end{proposition}

The case of constant  $ H_t, \Gamma_t $, but time-dependent  $ f_t $ naturally arises in physical applications for the semiclassical laser field with varying amplitude. In such a case the evolution of the second moments is defined by the previous proposition and the evolution of mean could be defined by the following proposition.

\begin{proposition}
	If the functions $ H_t, \Gamma_t $ in \eqref{eq:firstMoment} are constants $ H, \Gamma$ and $ f_t $ is a continuous function, then the solution of Eq.~\eqref{eq:firstMoment} takes the form
	\begin{equation}\label{ftSol}
	m_t = e^{J \left(i H + \frac{\Gamma^T - \Gamma}{2}\right) t} m_0 + i \int_0^t e^{J \left(i H + \frac{\Gamma^T - \Gamma}{2}\right) (t -t')}  J f_{t'} dt'.
	\end{equation}
	If, in addition, the function $ f_t $ has the form $ f_t = \sum_{k=1}^N \check{f}_k e^{i \omega_k t} $ and $ \det \left( J \left(i H + \frac{\Gamma^T - \Gamma}{2}\right) - i \omega_k I \right) \neq 0, \forall k $ (the absence of resonances), then \eqref{ftSol} takes the form
	\begin{equation*}
	m_t = e^{J \left(i H + \frac{\Gamma^T - \Gamma}{2}\right) t} m_0 + i \sum_k \frac{ e^{J \left(i H + \frac{\Gamma^T - \Gamma}{2}\right) t} - e^{i \omega_k t}I}{J \left(i H + \frac{\Gamma^T - \Gamma}{2}\right) - i \omega_k I} J  \check{f}_k.
	\end{equation*}
\end{proposition}

The  evolution map for Eq.~\eqref{eq:CauchyProbTime} at fixed time is closely related to the bosonic quantum Gaussian channels. They could be defined by the triple  $(G, l, \alpha)$ defining the transformation of the vector of means and the covariance matrix  (see Ref.~\cite{Holevo15}, p.~299):
\begin{equation*}
m' = G m + l, \qquad C' = G C G^T + \alpha.
\end{equation*}
\begin{proposition}
	\label{prop:QuantChan}
	The Gaussian channel defined by the map $e^{\mathcal{L}_{H, \Gamma, f} t}$ in the case of the non-degenerate matrix $ i H + \frac{\Gamma^T - \Gamma}{2} $ is characterized by the triple $ (G_t, l_t, \alpha_t) $
	\begin{align*}
	G_t  = e^{ J \left(i H + \frac{\Gamma^T - \Gamma}{2}\right) t}, \quad l_t=  \frac{ e^{J \left(i H + \frac{\Gamma^T - \Gamma}{2}\right) t} - I}{J \left(i H + \frac{\Gamma^T - \Gamma}{2}\right)} J i f,
	\\
	\alpha_t = \int_0^t e^{ J \left(i H + \frac{\Gamma^T - \Gamma}{2}\right) (t - t')}  J \frac{\Gamma + \Gamma^T}{2}J  e^{\left(-i H + \frac{\Gamma^T - \Gamma}{2}\right) J (t - t')} d t'.
	\end{align*}
\end{proposition}

\section{Fermionic case}\label{sec:femionic}

We consider the Hilbert space $\otimes_{j=1}^n\mathbb{C}^2 = \mathbb{C}^{2^n}$. It is possible to introduce $n$ pairs of fermionic creation and annihilation operators (see, for example, Ref.~\cite{Tak11}, p.~309 or Ref.~\cite{combescure2012coherent}, p.~312) by the formulae
\begin{align*}
c_i = (\otimes_{j=1}^{i-1} \sigma_3)\otimes \sigma^- \otimes (\otimes_{j=i+1}^{n} I_2),\qquad i =1, \ldots, n,
\\
c_i^{\dagger} = (\otimes_{j=1}^{i-1} \sigma_3)\otimes \sigma^+ \otimes (\otimes_{j=i+1}^{n} I_2),\qquad i =1, \ldots, n,
\end{align*}
where $\sigma_3, \sigma^-, \sigma^+$  are Pauli matrices and $I_2$ is the identity matrix
\begin{equation*}
\sigma_3 = \begin{pmatrix}
1 & 0\\
0 & -1
\end{pmatrix},
\quad
\sigma^- =  \begin{pmatrix}
0 & 0\\
1 & 0
\end{pmatrix},
\quad
\sigma^+ =  \begin{pmatrix}
0 & 1\\
0 & 0
\end{pmatrix},
\quad
I_2 =  \begin{pmatrix}
1 & 0\\
0 & 1
\end{pmatrix}.
\end{equation*}
The fermionic creation and annihilation operators satisfy the canonical anticommutation relations (CARs)
\begin{equation*}
\{c_i^{\dagger}, c_j \} = \delta_{ij}, \qquad \{c_i, c_j\} = 0.
\end{equation*}
Let us define the $2n$-dimensional vector  $\mathfrak{c} = (c_1, \ldots, c_n, c_1^{\dagger}, \ldots, c_n^{\dagger})^T$ of annihilation and creation operators and
\begin{align*}
f^T \mathfrak{c} &\equiv  \sum_{i=1}^n( f_i c_i + f_{i+n-1} c_i^{\dagger} ), \qquad \forall f \in \mathbb{C}^{2n},\\
\mathfrak{c}^T K \mathfrak{c} &\equiv \sum_{i=1}^n \sum_{j=1}^n( K_{i,j} c_i c_j + K_{i+n,j} c_i^{\dagger} c_j + K_{i,j+n} c_i c_j^{\dagger} + K_{i+n,j+n} c_i^{\dagger} c_j^{\dagger}), \; \forall K \in \mathbb{C}^{2n \times 2n}.
\end{align*}
Then CARs take the form
\begin{equation}\label{anti_comm}
\{f^T\mathfrak{c}, g^T \mathfrak{c} \} = f^T Eg, \qquad \forall f, g \in \mathbb{C}^{2n},
\end{equation}
where the matrix $E$ is defined by \eqref{JEIdef}. One could also represent it in the form
\begin{equation}\label{f_comm}
[f^T \mathfrak{c} , g^T\mathfrak{c} ] = 2 f^T \mathfrak{c}  g^T\mathfrak{c} - f^T Eg
\end{equation}
which is useful for computation of commutators.

We preserve definition \ref{tildeDef} for the $\sim$-conjugation but in the fermionic case the Hermitian conjugation of linear and quadratic forms could be calculated by the formulae $(g^T \mathfrak{c})^{\dagger} = \tilde{g}^T \mathfrak{c}$  and $(\mathfrak{c}^T K \mathfrak{c})^{\dagger} = - \mathfrak{c}^T \tilde{K} \mathfrak{c}$, where $K = - K^T$. Analogously to the bosonic case we use $ \mathrm{tr} \; $ for the trace with respect to $\otimes_{j=1}^n\mathbb{C}^2$ and $ \mathrm{Tr} \; $ for the trace with respect to $ \mathbb{C}^{2n} $. We also call a self-adjoint non-negatively defined matrix in $ \mathbb{C}^{2^n} $ with trace $ 1 $ \textit{a density matrix}.

\subsection{Basic calculations with quadratic and linear forms}
\label{subsec:basicFerm}

One could antisymmetrize a quadratic form by the formula
\begin{equation*}
\mathfrak{c}^T A \mathfrak{c} = \mathfrak{c}^T \frac{A - A^T}{2} \mathfrak{c} + \frac12 \mathrm{Tr} \; AE,
\end{equation*}
which allows one to use only antisymmetric forms in calculations.

The proofs of the following lemmas could be found in Ref.~\cite{Teretenkov17b}.

\begin{lemma}
	\label{f:commLemma}
	(Commutation relations for linear and quadratic forms)
	Let $A = -A^T, B= -B^T \in \mathbb{C}^{2 n \times 2n}, g,f \in \mathbb{C}^{2 n}$,
	then
	\begin{align*}
	\left[\frac12 \mathfrak{c}^T A \mathfrak{c} + f^T \mathfrak{c}, \frac12\mathfrak{c}^T B \mathfrak{c} + g^T \mathfrak{c}\right] = \frac12 \mathfrak{c}^T (A E B - B E A)\mathfrak{c} &+ \mathfrak{c}^T  \left(A E g - B E f \right) + \nonumber
	\\
	& + \mathfrak{c}^T ( f g^T - g f^T) \mathfrak{c}.
	\end{align*}
\end{lemma}

\begin{lemma}(Quadratic and linear forms sandwiched by exponentials.)
	\label{lemma_exp}
	Let $A=-A^T, B=-B^T \in \mathbb{C}^{2n \times 2n}$ and $f \in \mathbb{C}^{2n}$, then
	\begin{equation*}
	e^{ \frac12 \mathfrak{c}^T A \mathfrak{c}}\left( \frac12 \mathfrak{c}^T B \mathfrak{c} + g^T \mathfrak{c} \right) e^{- \frac12 \mathfrak{c}^T A \mathfrak{c}} = \frac12 \mathfrak{c}^T e^{A E}B e^{-E A} \mathfrak{c} + g^T e^{-E A} \mathfrak{c}.
	\end{equation*}
\end{lemma}

\begin{lemma}(Derivative of exponential.)
	Let $K_t = K_t^T \in \mathbb{C}^{2n \times 2n} $ be a differentiable function with respect to $ t $ for $ t \in \mathbb{R}_+ $, then
	\begin{equation*}
	\left(\frac{d}{dt} e^{ \frac12 \mathfrak{c}^T K_t \mathfrak{c}} \right) e^{ -\frac12 \mathfrak{c}^T K_t \mathfrak{c}} = \frac12 \mathfrak{c}^T \left(\frac{d}{dt} e^{K_t E} \right) e^{-K_t E} E\mathfrak{c},
	\end{equation*}
	and the matrix $\left(\frac{d}{dt} e^{K_t E} \right) e^{-K_t E} E$ is antisymmetric.
\end{lemma}

Similar to lemma \ref{lemma3} this result is also based on the Feynman-Wilcox formula\cite{Chebotarev12}.

\subsection{Quadratic superoperators}
\label{subsec:QuadFerm}

In this subsection we define the GKSL generators which are quadratic in fermionic creation and annihilation operators.

\begin{definition}
	We call the linear operator $\mathcal{L}: \mathbb{C}^{2^n \times 2^n} \rightarrow \mathbb{C}^{2^n \times 2^n} $ of the form
	\begin{equation}\label{f_quad_gen}
	\mathcal{L}(\rho) = \mathfrak{c}^T \Gamma \rho \mathfrak{c} + \mathfrak{c}^T \Gamma_L  \mathfrak{c} \rho + \rho \mathfrak{c}^T \Gamma_R  \mathfrak{c} + \rho g_R^T \mathfrak{c} + g_L^T \mathfrak{c} \rho + \lambda \rho,
	\end{equation}
	where $\Gamma, \Gamma_L = -\Gamma_L^T, \Gamma_R = -\Gamma_R^T\in \mathbb{C}^{2n \times 2n}, g_L, g_R \in \mathbb{C}^{2n}, \lambda \in \mathbb{C}$, \textit{a quadratic fermionic superoperator}.
\end{definition}

\begin{definition}
	Let $ \mathcal{L}: \mathbb{C}^{2^n \times 2^n} \rightarrow \mathbb{C}^{2^n \times 2^n} $ be a quadratic superoperator of form \eqref{f_quad_gen}. Let us define a superoperator $\mathcal{L}^*: \mathbb{C}^{2^n \times 2^n} \rightarrow \mathbb{C}^{2^n \times 2^n} $ by the formula
	\begin{equation*}
	\mathrm{tr} \; ((\mathcal{L}^*(\hat{X}))^{\dagger} \rho ) = \mathrm{tr} \; (\hat{X}^{\dagger} \mathcal{L}(\rho)).
	\end{equation*}
\end{definition}

\begin{definition}
	Let $ \mathcal{L}: \mathbb{C}^{2^n \times 2^n} \rightarrow \mathbb{C}^{2^n \times 2^n} $  be a quadratic superoperator of form \eqref{f_quad_gen}. Let us define a superoperator $\mathcal{L}^+: \mathbb{C}^{2^n \times 2^n} \rightarrow \mathbb{C}^{2^n \times 2^n} $ by the formula
	\begin{equation*}
	\mathcal{L}^+(\rho) = \left( \mathcal{L}(\rho^{\dagger}) \right)^{\dagger}.
	\end{equation*}
\end{definition}

\begin{lemma}\label{f_plus_conj}
	Let $\mathcal{L}$ be a quadratic superoperator of form \eqref{f_quad_gen}, where
	\begin{align*}
	\mathcal{L}^+(\rho) &= \mathfrak{c}^T \tilde{\Gamma}^T \rho \mathfrak{c} - \mathfrak{c}^T \tilde{\Gamma}_R  \mathfrak{c} \rho - \rho \mathfrak{c}^T \tilde{\Gamma}_L \mathfrak{c} + \rho \tilde{g}_L^T \mathfrak{c} + \tilde{g}_R^T \mathfrak{c} \rho +\overline{ \lambda} \rho,
	\\
	\mathcal{L}^*(\hat{X}) &= \mathfrak{c}^T \tilde{\Gamma} \hat{X} \mathfrak{c} - \mathfrak{c}^T \tilde{\Gamma}_L  \mathfrak{c} \hat{X} - \hat{X} \mathfrak{c}^T \tilde{\Gamma}_R \mathfrak{c} + \hat{X} \tilde{g}_R^T \mathfrak{c} + \tilde{g}_L^T \mathfrak{c} \hat{X} +\overline{ \lambda} \hat{X}.
	\end{align*}
\end{lemma}

\begin{lemma}
	\label{lemma:derf}
	The superoperator $\mathcal{L}$ of form \eqref{f_quad_gen} satisfies the conditions $\mathcal{L}^+ = \mathcal{L}$ and $\mathcal{L}^*(I) = 0$ if and only if
	\begin{equation}\label{f:pos_gen}
	\mathcal{L}(\rho) = \mathcal{L}_{H, \Gamma, f}(\rho) \equiv -i\left[\frac{1}{2} \mathfrak{c}^T H \mathfrak{c} + f^T \mathfrak{c} , \rho \right] +  \mathfrak{c}^T \rho \Gamma \mathfrak{c}  - \frac{1}{2} \mathfrak{c}^T \Gamma^T \mathfrak{c} \rho  - \frac{1}{2} \rho \; \mathfrak{c}^T \Gamma^T \mathfrak{c} ,
	\end{equation}
	where $ H = -H^T = -\tilde{H} $,  $ \Gamma^T = \tilde{\Gamma} $.
\end{lemma}

\begin{lemma}\label{f:lemmaLindbladform}
	If $\hat{H} = \frac12 \mathfrak{c}^T H \mathfrak{c} + h^T \mathfrak{c} $ ($ H =  H^T, H = - \tilde{H} $ , $ h = \tilde{h} $) and $\hat{C}^{(i)}= \gamma_i^T \mathfrak{c}$, then the GKSL generator \eqref{lind_gen1}  takes the form \eqref{f:pos_gen}, where
	\begin{equation}\label{f:lind_param}
	\Gamma = \sum\limits_i \gamma_i \tilde{\gamma}_i^T,  \qquad l = \sum\limits_i \overline{c_i} \gamma_i, \qquad f = h + i \; \frac{l - \tilde{l}}{2}.
	\end{equation}
	Moreover, it is possible to represent $\Gamma$ and $ f $ in form \eqref{f:lind_param} if and only if
	\begin{equation*}
	\Gamma^T = \tilde{\Gamma}, \quad \Gamma E \geqslant 0, \quad f = \tilde{f}.
	\end{equation*}
\end{lemma}

\begin{lemma}\label{lemma:adjointSuperOpf}
	The adjoint superoperator for \eqref{f:pos_gen} has the form
	\begin{equation*}
	\mathcal{L}_{H, \Gamma, f}^* (\hat{X}) = i\left[\frac{1}{2} \mathfrak{c}^T H \mathfrak{c} + f^T \mathfrak{c} , \hat{X} \right] +  \mathfrak{c}^T \hat{X} \Gamma^T \mathfrak{c}  - \frac{1}{2} \mathfrak{c}^T \Gamma^T \mathfrak{c} \hat{X}  - \frac{1}{2} \hat{X} \; \mathfrak{c}^T \Gamma^T \mathfrak{c},  \hat{X} \in \mathbb{C}^{2^n \times 2^n}  .
	\end{equation*}
\end{lemma}

\subsection{Gaussian states}
\label{subsec:GaussFerm}

In this subsection we define the even fermionic Gaussian state. The main distinction from the bosonic case consists in the fact that we consider only even Gaussian state, i.e. only quadratic terms without linear ones participate in the exponentials defining the states. From the physical point of view it is a result of superselection rules.\cite{Amosov17}

\begin{definition}
	We define \textit{the matrix of the second moments} $ D $ for the density matrix $ \rho $ by the formula
	\begin{equation*}
	D = \mathrm{tr} \; (\mathfrak{c} \; \mathfrak{c}^T \rho).
	\end{equation*}
\end{definition}

\begin{definition}
	We call the antisymmetric part of the matrix $ D $ of the second moments
	\begin{equation*}
	A = \frac{D - D^T}{2}
	\end{equation*}
	\textit{the anticovariance matrix}.	
\end{definition}

Sometimes in literature it is called the covariance matrix for fermionic states\cite{Grepl13}, but it is inconsistent with general definition of the covariance matrix (see  Ref.~\cite{Holevo15}, p.~21), so we prefer the term above. The symmetric part of $ D $ is defined by CARs \eqref{anti_comm} as $ D +D^T =E $, then $ D = A + E/2 $.

One could also define the characteristic function, but in such a case one has to extend the CARs algebra to the CARs-Grassmann algebra with generators anticommuting with each other and with generators of the CARs algebra (see Ref.~\cite{combescure2012coherent}, Subsec.~11.5.1):
\begin{equation*}
\{\theta_i, \theta_j\} = 0, \{\theta_i, \overline{\theta}_j\} = 0,  \{\theta_i, c_j\} = 0 ,  \{\theta_i, c_j^{\dagger}\} = 0,  \{\overline{\theta}_i, c_j\} = 0 ,  \{\overline{\theta}_i, c_j^{\dagger}\} = 0,
\end{equation*}
where $ i,j=1,\ldots,n $. The characteristic function is a Grassmann variable function. It could be considered as a function in a Banach space\cite{VlaVol84,VlaVol84a}, but it could also be considered as an element of the Grassmann algebra rather than a true function (see Ref.~\cite{Tak11}, Ch.~7, \S~2). The latter (pure algebraic) approach is sufficient for our purposes.
\begin{definition}
	We define the \textit{characteristic function} by the formula
	\begin{equation*}
	\chi_W (\bm{\theta}) \equiv \mathrm{tr} \; (\rho \; e^{\bm{\theta}^T E \mathfrak{c}}),
	\end{equation*}
	where $\bm{\theta} = (\theta_1, \ldots, \theta_n, \overline{\theta}_1, \ldots, \overline{\theta}_n)^T$.
\end{definition}

In the algebraic approach the trace here could be regarded as partial trace with respect to the $ \mathbb{C}^{2^n} $ for matrix representation of the CARs-Grassman algebra. Now one could define even Gaussian states similar to definition \ref{eq:charW}.

\begin{definition}
	\label{oddGaussSt}
	The state $ \rho $ is  called an \textit{even Gaussian state} if its characteristic function has the form
	\begin{equation*}
	\chi_W (\bm{\theta}) \equiv \mathrm{tr} \; (\rho \; e^{\bm{\theta}^T E \mathfrak{c}}) = e^{\frac12 \bm{\theta}^T A \bm{\theta} }.
	\end{equation*}
\end{definition}

\begin{proposition}
	The matrix $ D \in \mathbb{C}^{2n \times 2n} $ satisfies the conditions
	\begin{equation}
	\label{f:secMomMatrixCond}
	(D E)^+ = D E \geqslant 0, \qquad D + D^T = E
	\end{equation}
	if and only if there exists a (unique) even Gaussian state with the matrix of the second moments $ D $.
\end{proposition}

As in the bosonic case the conditions \eqref{f:secMomMatrixCond} are satisfied for the arbitrary density matrix.

Normally and anti-normally ordered characteristic functions could be also defined (see Ref.~\cite{Cahill99}, Sec.~VIII):
\begin{align*}
\chi_N (\bm{\theta}) &\equiv \mathrm{tr} \; (\rho \; e^{ \bm{\theta}^T \frac{E -J}{2} \mathfrak{c}} e^{\bm{\theta}^T \frac{E + J}{2} \mathfrak{c}}) = e^{\frac12 \bm{\theta}^T \left( A + \frac12 J \right)  \bm{\theta} },\\
\chi_A (\bm{\theta}) &\equiv \mathrm{tr} \; (\rho \; e^{i \bm{\theta}^T \frac{E + J}{2} \mathfrak{c}} e^{i \bm{\theta}^T \frac{E - J}{2} \mathfrak{c}}) =  e^{ \frac12 \bm{\theta}^T \left( A - \frac12 J \right) \bm{\theta}}.
\end{align*}

Non-pure Gaussian states could be also represented in the exponential form
\begin{equation}\label{eq:statef}
\rho = e^{\frac12 \mathfrak{c}^T K \mathfrak{c} + s},
\end{equation}
where $ K = - \tilde{K} \in \mathbb{C}^{2n \times 2n} $ and $ s \in \mathbb{R} $ is defined by
\begin{equation}\label{tr_norm}
e^{s}  = \left(\sqrt{\det(e^{KE} + I)}\right)^{-1}.
\end{equation}
Formula \eqref{tr_norm} is based on the following result (see Ref.~\cite{DodMan83}, p.~154 or Ref.~\cite{Grepl13}, p.~15).
\begin{proposition}
	The matrix $K = - K^T= - \tilde{K} \in \mathbb{C}^{2n \times 2n}$ could be bidiagonalized
	\begin{equation*}
	K = O \begin{pmatrix}
	0 & -\Lambda\\
	\Lambda & 0
	\end{pmatrix} O^T
	\end{equation*}
	by orthogonal transform $O E O^T = O, \tilde{O} = O$.
\end{proposition}

\begin{proposition}
	Let $K = - K^T= - \tilde{K} \in \mathbb{C}^{2n \times 2n}$, then there exists the Gaussian state \eqref{eq:statef} with the matrix of the second moments $ D $ defined by the formula
	\begin{equation*}
	D =   E  (I + e^{KE})^{-1}.
	\end{equation*}
\end{proposition}

Let us denote the normal symbol (see Ref.~\cite{Ber86}, Ch.~II, Sec.~5) of the density matrix $\rho$ by $\rho (\bm{\theta})$ and connect it with the parameters  in exponential \eqref{eq:statef} by the following proposition. Its proof could be found in Ref.~\cite{Teretenkov17b}.
\begin{proposition}\label{f:norm_symb}
	Let $K = -K^T = - \tilde{K} \in \mathbb{C}^{2n \times 2n}$, and $s$ be defined by the normalization condition, i.e. by formula \eqref{tr_norm}, then the normal symbol is defined by the formula $\rho (\bm{\theta}) =  e^{\frac12 \bm{\theta}^T \; R \; \bm{\theta} + r} $, where
	\begin{equation*}
	R = 2 \left( J + E \frac{e^{K E} + I}{e^{K E} - I} \right)^{-1}, \qquad e^r = \frac{1}{\sqrt{\det R}}.
	\end{equation*}
\end{proposition}

\subsection{Dynamics of moments and Gaussian solutions}
\label{subsec:DynFerm}

In this subsection we obtain the Gaussian solutions of the equation
\begin{equation}\label{eq:CauchyProbTimef}
\frac{d}{dt}\rho_t = \mathcal{L}_{H_t, \Gamma_t, 0}(\rho_t),
\end{equation}
i.e. we assume $ f = 0 $. From the physical point of view the equations of such a type only are considered because the evolution preserves the even Gaussian states in such a case, but we have to mention that there is a mathematical reason for this. Namely, for $ f = 0 $ we obtain the following lemma which is fully similar to lemma \ref{conjGenAction} for the bosonic case.

\begin{lemma}
	\label{lemma:adjActf}
	The action of the adjoint generator on the pair-wise products of creation and annihilation operators is defined by the formula
	\begin{equation*}
	\mathcal{L}_{H, \Gamma, 0}^* \left( \mathfrak{c} \; \mathfrak{c}^T \right) =  E \left( - i H - \frac{\Gamma^T + \Gamma}{2} \right) \mathfrak{c} \; \mathfrak{c}^T + \mathfrak{c} \; \mathfrak{c}^T \left(i H - \frac{\Gamma^T + \Gamma}{2} \right) E + E \Gamma^T E.
	\end{equation*}
\end{lemma}

For $ f \neq 0 $ the formula for $ \mathcal{L}_{H, \Gamma, f}^*  ( \mathfrak{c} \; \mathfrak{c}^T ) $ contains the terms which are cubic in $ \mathfrak{c} $ and, hence, does not allow one to close the equation for the second moments. Lemma \ref{lemma:adjActf} follows immediately from lemma \ref{lemma:adjointSuperOpf}.

\begin{theorem} Let the density matrix $ \rho_t $ satisfy Eq.~\eqref{eq:CauchyProbTimef}, then the matrix of the second moments satisfies
	\begin{equation}\label{eq:secondMomentsf}
	\frac{d}{dt} D_t   =  E \left( - i H_t - \frac{\Gamma_t^T + \Gamma_t}{2} \right) D_t + D_t \left(i H_t - \frac{\Gamma_t^T + \Gamma_t}{2} \right) E + E \Gamma_t^T E.
	\end{equation}
\end{theorem}

Similar to the bosonic case it could be proved by substitution of \eqref{eq:statef} into \eqref{eq:CauchyProbTimef} and applying lemmas \ref{f:commLemma}--\ref{lemma:derf} for Gaussian states\cite{Teretenkov17b} or by averaging the formula from lemma \ref{lemma:adjActf}\cite{Teretenkov19}.

\begin{proposition}\label{f:prop:covMatEq}
	Let the matrix $ D_t $ satisfy Eq.~\eqref{eq:secondMomentsf}, then the anticovariance matrix $A_t =(A_t + A_t^T)/2$ satisfies the equation
	\begin{equation}\label{eq:covMatf}
	\frac{d}{dt} A_t  =  E \left( - i H_t - \frac{\Gamma_t^T + \Gamma_t}{2} \right) A_t  + A_t \left(i H_t - \frac{\Gamma_t^T + \Gamma_t}{2} \right) E + E \frac{\Gamma_t^T - \Gamma_t}{2} E.
	\end{equation}
	If, in addition, the matrix $ D_0 $ satisfies the condition $D_0 + D_0^T = E$, then $D_t + D_t^T = E$ for all $ t>0 $.
\end{proposition}

As for the bosonic case several representations of Eqs. \eqref{eq:secondMomentsf}, \eqref{eq:covMatf} could be obtained (See Ref.~\cite{Teretenkov17b} for proof and discussion).

\begin{proposition}
	\label{f:prop:2nSol}
	Let $ G_t $ be a solution of the Cauchy problem
	\begin{equation*}
	\dot{G}_t = - E \left(i H_t +\frac{\Gamma_t^T + \Gamma_t}{2}\right) G_t, \qquad G_0 = I.
	\end{equation*}
	Then the solutions of Eqs.~\eqref{eq:secondMomentsf}, \eqref{eq:covMatf} could be represented in the form
	\begin{align*}
	D_t &= G_t D_0 G_t^T + \int_0^t G_{t} G_{t'}^{-1} E \Gamma_{t'}^T E (G_{t} G_{t'}^{-1})^T dt',\\
	A_t &= G_t A_0 G_t^T +  \int_0^t G_{t} G_{t'}^{-1} E \frac{\Gamma_{t'}^T - \Gamma_{t'}}{2} E (G_{t} G_{t'}^{-1})^T dt'.
	\end{align*}
\end{proposition}

\begin{proposition}
	\label{f:doubleEqProp}	
	The solution of Eq.~\eqref{eq:secondMomentsf} could be represented in the form  $D_t = X_t Y_t^{-1}$, where the matrices $ X_t $ and $ Y_t $ are defined by the following differential equation
	\begin{equation*}
	\frac{d}{dt}
	\begin{pmatrix}
	X_t\\
	Y_t
	\end{pmatrix}
	=
	\begin{pmatrix}
	E \left( - i H_t - \frac{\Gamma_t^T + \Gamma_t}{2} \right) &  E \Gamma_t^T E      \\
	0                              &  \left(-i H_t + \frac{\Gamma_t^T + \Gamma_t}{2} \right) E
	\end{pmatrix}
	\begin{pmatrix}
	X_t\\
	Y_t
	\end{pmatrix}
	\end{equation*}
	with an arbitrary initial condition such that $ X_0   Y_0^{-1} =  D_0$. In particular, one could assume $ X_0 = D_0, Y_0 = I $.
	
	Similarly, the solution of Eq.~\eqref{eq:covMatf} could be represented in the form $A_t = X_t' Y_t^{-1}$,where the matrices $ X_t' $ and $ Y_t $ are defined by the following differential equation
	\begin{equation*}
	\frac{d}{dt}
	\begin{pmatrix}
	X_t'\\
	Y_t
	\end{pmatrix}
	=
	\begin{pmatrix}
	E \left( - i H_t - \frac{\Gamma_t^T + \Gamma_t}{2} \right) &  E  \frac{\Gamma_t^T - \Gamma_t}{2} E      \\
	0                              &  \left(-i H + \frac{\Gamma_t^T + \Gamma_t}{2} \right) E
	\end{pmatrix}
	\begin{pmatrix}
	X_t'\\
	Y_t
	\end{pmatrix}
	\end{equation*}
	with an arbitrary initial condition such that $ X_0'  Y_0^{-1} =  A_0$. In particular, one could assume $ X_0' = A_0, Y_0 = I $.
\end{proposition}

\begin{proposition}
	If the functions $ H_t, \Gamma_t $ in \eqref{eq:secondMomentsf} are constants $ H, \Gamma $, then the solution of Eq.~\eqref{eq:secondMomentsf} takes the form
	\begin{equation*}
	D_t = e^{ E \left( - i H - \frac{\Gamma^T + \Gamma}{2} \right) t} D_0 e^{\left(i H - \frac{\Gamma^T + \Gamma}{2} \right) E t} +   \int_0^t e^{ E \left( - i H - \frac{\Gamma^T + \Gamma}{2} \right) t'}  E \Gamma^T E  e^{\left(i H - \frac{\Gamma^T + \Gamma}{2} \right) E t'} d t'.
	\end{equation*}
	If the functions $ H_t, \Gamma_t $ in \eqref{eq:covMatf} are constants $ H, \Gamma $, then the solution of Eq.~\eqref{eq:covMatf} takes the form
	\begin{equation*}
	A_t = e^{ E \left( - i H - \frac{\Gamma^T + \Gamma}{2} \right) t} A_0 e^{\left(i H - \frac{\Gamma^T + \Gamma}{2} \right) E t} +  \int_0^t e^{ E \left( - i H - \frac{\Gamma^T + \Gamma}{2} \right) t'}  E \frac{\Gamma^T - \Gamma}{2}E  e^{\left(i H - \frac{\Gamma^T + \Gamma}{2} \right) E t'} d t'.
	\end{equation*}
\end{proposition}

The parity-preserving Gaussian fermionic channel\cite{Grepl13} is defined by the pair $ (G, \alpha) $ leading to the following transform of the second moments
\begin{equation*}
A' = G A G^T + \alpha.
\end{equation*}

\begin{proposition}
	\label{prop:QuantChanf}
	The Gaussian channel defined by the map $e^{\mathcal{L}_{H, \Gamma, 0} t}$ in the case of the non-degenerate matrix $ i H - \frac{\Gamma^T + \Gamma}{2} $ is characterized by the pair $ (G_t, \alpha_t) $
	\begin{align*}
	G_t  = e^{ E \left( - i H - \frac{\Gamma^T + \Gamma}{2} \right) t}, \;
	\alpha_t = \int_0^t e^{ E \left( - i H - \frac{\Gamma^T + \Gamma}{2} \right) t'}  E \frac{\Gamma^T - \Gamma}{2}E  e^{\left(i H - \frac{\Gamma^T + \Gamma}{2} \right) E t'} d t'.
	\end{align*}
\end{proposition}

\section{Conclusion}\label{concl}

We have presented the solutions of GKSL equations with generators which are quadratic in bosonic or fermionic creation and annihilation operators. The solutions are expressed in terms of equation for the second and the first moments which are presented in several forms. The solutions with constant coefficients are implemented in our package. Its source code could be found in Ref.~\cite{quadratica2017}.

One of possible directions of further development is the generalization of the above results to the different models of non-Markovian evolution which are widely discussed now (see Refs.~\cite{maniscalco2006non}--\cite{semin2016minimalistic}). The non-GKSL time-dependent generators could appear in such a case (see Ref.~\cite{Bro10}, Ch.~10). In particular, the time-convolutionless master equation approach leads to such generators (see Ref.~\cite{breuer2001time} or Ref.~\cite{Bro10}, Ch.~9). Non-GKSL evolution of a one-dimensional quantum Brownian particle was considered in Refs.~\cite{maniscalco2004lindblad} and \cite{maniscalco2009non}. Non-GKSL generators lead to positive but non-completely positive maps which are also actively discussed\cite{shaji2005s,man2005partial,DePalma15,filippov2012local,maniscalco2007entanglement}. The master equations with time-non-local generators and their relations to the time-local equations are also intensively studied now\cite{chruscinski2010memory, chruscinski2010non, chruscinski2014time,siudzinska2017memory}.

Unitary symmetries for GKSL generators were widely studied\cite{Holevo93, Holevo95, Holevo96}. So another direction for further study is understanding analogs for non-unitary conservation laws for the GKSL case, which is inspired by proposition \ref{prop:commut}.

\section*{Acknowledgment}
The author wishes to express gratitude to A.\,M. Chebotarev, A.\,S. Holevo, V.\,I. Man'ko, A.\,M.~Basharov, I.\,V.~Volovich, S.\,N. Filippov and E.\,Greplova for fruitful discussion of the problems considered in the work. I would also like to thank the referees for their valuable comments.


\begin{thebibliography}{999}
	%
	%
	%
	%
	%
	
	\bibitem{Fried1953}
	K.\,O.~Friedrichs, Mathematical aspects of the quantum theory of fields. Part V. Fields modified by linear homogeneous forces,
	{\small\it Comm. on Pure and App. Math.}
	{\small\bf 6}(1) (1953) 1--72 .
	
	\bibitem{Ber86}
	F.\,A. Berezin,
	{\small\it Metod vtorichnogo kvantovaniya}
	(Nauka, M., 1986)
	
	\bibitem{Manko79}
	I.\,A.~Malkin, V.\,I.~Man'ko,
	\textit{Dinamicheskie simmetrii i kogerentnye sostoyaniya kvantovyh sistem}
	(Nauka, M., 1979)
	
	\bibitem{Manko87}
	V.\,V.~Dodonov, V.\,I.~Man'ko,
	Invarianty i evolyutsiya nestatsionarnykh kvantovykh sistem,
	Tr. FIAN, \textbf{183}  (Nauka, Moscow, 1987)
	
	\bibitem{dodonov2003theory}
	V.\,V.~Dodonov, V.\,I.~Man'ko,
	{\small\it Theory of nonclassical states of light}
	(Taylor and Francis, London-New York, 2003)
	
	\bibitem{dodonov2002nonclassical}
	V.\,V.~Dodonov,
	'Nonclassical' states in quantum optics: a 'squeezed' review of the first 75 years,
	{\small\it J. of Opt. B}
	{\small\bf 4}(1) (2002) R1--R33 .
	
	\bibitem{Cheb12}
	A.\,M.~Chebotarev, T.\,V.~Tlyachev, A.\,A.~Radionov,
	Generalized Squeezed States and Multimode Factorization Formula,
	{\small\it Math. Notes}
	{\small\bf 92}(5) (2012) 700--713.
	
	\bibitem{Cheb11}
	A.\,M.~Chebotarev, T.\,V.~Tlyachev, A.\,A.~Radionov,
	Squeezed States and Their Applications to Quantum Evolution,
	{\small\it Math. Notes}
	{\small\bf 89}(4) (2011) 577--595.
	
	\bibitem{chebotarev2014normal}
	A.\,M.~Chebotarev, T.\,V.~Tlyachev,
	Normal forms, inner products, and Maslov indices of general multimode squeezings,
	{\small\it Math. Notes}
	{\small\bf 95}(5--6) (2014) 721--737.
	
	\bibitem{achmanov1964}
	S.\,A.~Achmanov and R.\,V.~Khokhlov,
	{\small\it Problemy nelineynoy optici}
	[in Russian] (VINITI, Moscow, 1964)
	
	\bibitem{scalli2003}
	M.\,O.~Scully, M.\,S.~Zubairy,
	{\small\it Quantum optics}
	(Cambridge University Press, Cambridge, 1997).
	
	
	\bibitem{maimistov2013nonlinear}
	A.\,I.~Maimistov and A.\,M.~Basharov,
	{\small\it Nonlinear optical waves}
	(Dordrecht Kluwer Acad. Publ.,  London, 1999).
	
	\bibitem{dodonov1995even}
	V~V Dodonov, V~I Man'ko, and D~E Nikonov,
	Even and odd coherent states for multimode parametric systems,
	{\small\it Phys. Rev. A}
	{\small\bf 51}(4) (1995) 3328.
	
	\bibitem{chirkin2007statistic}
	A.\,S.~Chirkin and M.\,Yu.~Saigin,
	Statistic and information characterization of tripartite entangled states,
	{\small\it J. of Russ. Laser Research}
	{\small\bf 28}(5) (2007) 505--515.
	
	\bibitem{Chirkin2007}
	A.\,S.~Chirkin, I.\,V.~Shutov,
	On the possibility of the nondegenerate parametric amplification of optical waves at low-frequency pumping,
	{\small\it JETP Lett.}
	{\small\bf 86}(11) (2008) 693--697.
	
	\bibitem{chirkin2009four}
	A.\,S.~Chirkin, M.\,Yu~Saigin,
	Four-mode entangled states in coupled nonlinear optical processes and teleportation of two-mode entangled CV state,
	{\small\it Phys. Scr.}
	{\small\bf 135} (2009) 014029.
	
	\bibitem{saygin2010simultaneous}
	M.\,Yu~Saigin, A.\,S.~Chirkin,
	Simultaneous parametric generation and up-conversion of entangled optical images,
	{\small\it JETP}
	{\small\bf 111}(1) (2010), 11--21.
	
	\bibitem{tlyachev2014canonical}
	T.\,V.~Tlyachev,  A.\,M.~Chebotarev, A.\,S.~Chirkin,
	Canonical transformations and multipartite coupled parametric processes,
	{\small\it Phys. Scr.}
	{\small\bf 160} (2014) 014041.
	
	\bibitem{tlyachev2013new}
	T.\,V.~Tlyachev,  A.\,M.~Chebotarev, A.\,S.~Chirkin,
	A new approach to quantum theory of multimode coupled parametric processes,
	{\small\it Phys. Scr.}
	{\small\bf 153} (2013) 014060.
	
	\bibitem{huang2009entangling}
	S.~Huang, G.\,S.~Agarwal,
	Entangling nanomechanical oscillators in a ring cavity by feeding squeezed light,
	{\small\it New J. of Phys.}
	{\small\bf 11}(10) (2009) 103044.
	
	\bibitem{huang2009squeezing}
	S.~Huang, G.\,S.~Agarwal,
	Squeezing of a nanomechanical oscillator,
	arXiv:0909.4234 (2009).
	
	\bibitem{Bogol1947}
	N.\,N.~Bogolubov,
	K teorrii svertekhuchesti
	{\small\it Isv. AN SSSR, Ser. fiz.}
	{\small\bf11}(1) (1947) 77--90 [in Russian].
	
	\bibitem{kvas2014v4}
	I.\,A.~Kvasnikov,
	{\small\it Termodinamika i statisticheskaya fizikf. T. 4: Kvantovaya statistica}
	(URSS, Moscow, 2014) [in Russian].
	
	\bibitem{chernikov1968system}
	N.\,A.~Chernikov,
	The system whose hamiltonian is a time-dependent quadratic,
	{\small\it JETP}
	{\small\bf 26}(3) (1968) 603--608.
	
	\bibitem{lewis1969exact}
	H.\,R.~Lewis Jr.,  W.\,B.~Riesenfeld,
	An exact quantum theory of the time-dependent harmonic oscillator and of a charged particle in a time-dependent electromagnetic field,
	{\small\it J. Math. Phys.}
	{\small\bf 10}(8) (1969) 1458--1473.
	
	\bibitem{Bud1987One}
	V.\,G.~Budanov,
	An equation for disentangling time-ordered exponentials with arbitrary quadratic generators,
	{\small\it Theoret. and Math. Phys.}
	{\small\bf 71}(3) (1987) 570--574.
	
	\bibitem{castanos2006squeezing}
	O.~Castanos, R.~Lopez-Pena, M.\,A.~Man'ko, and V.\.I.~Man'ko,
	Squeezing Operator and Squeeze Tomography,
	in {\small\it Topics In Mathematical Physics, General Relativity And Cosmology In Honor Of Jerzy Plebanski}
	(World Scientific, Singapore, 2006), pp. 109--120.
	
	\bibitem{williamson1936algebraic}
	J.~Williamson,
	On the algebraic problem concerning the normal forms of linear dynamical systems,
	{\small\it American J. of Math.}
	{\small\bf 58}(1) (1936) 141--163.
	
	\bibitem{williamson1940algebraic}
	J.~Williamson,
	An algebraic problem involving the involutory integrals of linear dynamical systems,
	{\small\it American J. of Math.}
	{\small\bf 62}(1) (1940) 881--911.
	
	\bibitem{williamson1937normal}
	J.~Williamson,
	On the normal forms of linear canonical transformations in dynamics,
	{\small\it American J. of Math.}
	{\small\bf 59}(3) (1937) 599--617.
	
	\bibitem{alekseev2009squeezed}
	P.\,S.~Alekseev and F.\,V.~Moroseev,
	Squeezed states in the semiclassical limit,
	{\small\it JETP}
	{\small\bf 108}(4) (2009) 571--582.
	
	\bibitem{dodonov2002nonclassical}
	V.\,V.~Dodonov,
	Nonclassical'states in quantum optics: a 'squeezed' review of the first 75 years,
	{\small\it J. Opt. B: Quant. Semiclass. Opt.}
	{\small\bf 4} (2002) R1--R33.
	
	\bibitem{DodMan83}
	V.\,V.~Dodonov, V\,I.~Man'ko,
	Integrals of Motion and Dynamics of General Non-Stationary Quadratic Fermi-Bose Systems,
	{\small\it Trudy FIAN}
	{\small\bf 153} (1983) 145--193 [in Russian].
	
	\bibitem{landau1927}
	L.\,D.~Landau,
	The Damping Problem in Wave Mechanics,
	{\small\it Z. Phys.}
	{\small\bf 45} (1927) 430-441 [in German].
	
	\bibitem{perelomov1987}
	A.\,Perelomov,
	{\small\it Generalized coherent states and their applications} (Springer-Verlag, Berlin-Heidelberg, 1986).
	
	
	\bibitem{Bogolubov1939}
	N.\,M.~Krylov and N.\,N.~Bogolubov,
	On derivation of Fokker-Planck equations by perturbation-theory method based on spectral properties of the perturbation Hamiltonian,
	{\small\it Zap. Kaf. Mat. Fiz. Ukrain. Akad. Nauk}
	{\small\bf 4} (1939) 5--80 [in Ukrainian].
	
	\bibitem{gardiner2004quantum}
	C.~Gardiner and P.~Zoller,
	{\small\it Quantum noise: a handbook of Markovian and non-Markovian quantum stochastic methods with applications to quantum optics}
	(Springer, Berlin-Heidelberg, 2004).
	
	\bibitem{Bro10}
	H.-P.~Breuer and F.~Petruccione,
	{\small\it The theory of open quantum systems}
	(Oxford University Press, Oxford, 2002).
	
	\bibitem{bausch1967description}
	R.~Bausch, A.~Stahl,
	On the description of noise in quantum systems,
	{\small\it Zeitschrift f{\"u}r Physik A}
	{\small\bf 204}(1) (1967) 32--46.
	
	\bibitem{gorini1976completely}
	V.~Gorini, A.~Kossakowski, and E.\,C.\,D.~Sudarshan,
	Completely positive dynamical semigroups of N-level systems,
	{\small\it J. Math. Phys.}
	{\small\bf 17}(5) (1976) 821--825.
	
	\bibitem{lindblad1976generators}
	G.~Lindblad,
	On the generators of quantum dynamical semigroups,
	{\small\it Comm. in Math. Phys.}
	{\small\bf 48}(2) (1976) 119--130.
	
	\bibitem{chruscinski2017brief}
	D.~Chruscinski and S.~Pascazio,
	A brief history of the GKLS equation,
	arXiv preprint arXiv:1710.05993 (2017).
	
	\bibitem{accardi2002lectures}
	L.~Accardi and S.~Kozyrev,
	Lectures on quantum interacting particle systemss, in
	{\small\it Quantum Interacting Particle Systems} (Trento, 2000), QP?PQ: Quantum Probab. White Noise Anal., Vol. 14 (World Scientific, 2002), pp. 1--195.
	
	\bibitem{accardi2013quantum}
	L.~Accardi, Y.\,G.~Lu, and I.~Volovich,
	{\small\it Quantum theory and its stochastic limit}
	(Springer-Verlag, Berlin-Heidelberg, 2002).
	
	\bibitem{Pech04}
	L.~Accardi, A.\,N.~Pechen, and I.\,V.~Volovich,
	Quantum stochastic equation for the low density limit,
	{\small\it J. of Phys. A: Math. and General}
	{\small\bf 35}(23) (2002) 4889.
	
	\bibitem{davies1976quantum}
	E.\,B.~Davies
	{\small\it Quantum Theory of Open Systems} (Academic Press, London, 1976).
	
	\bibitem{basharov2011spontaneous}
	A.\,M.~Basharov,
	Spontaneous emission of the non-Wiener type,
	{\small\it JETP}
	{\small\bf 113}(3) (2011) 376.
	
	
	\bibitem{Basharov2012}
	A.\,M.~Basharov,
	Quantum theory of open systems based on stochastic differential equations of generalized Langevin (non-Wiener) type. {\small\it JETP}
	{\small\bf 115}(3) (2012) 371--391.
	
	\bibitem{Basharov2014}
	A.\,M.~Basharov,
	A theory of open systems based on stochastic differential equations,
	{\small\it Opt. Spectrosc.}
	{\small\bf 116}(4) (2014) 495--503.
	
	\bibitem{Accardi17}
	L.~Accardi and Y.\,G.~Lu,
	The first 40 years of GKSL generators and some proposal for the future,
	{\small\it Open Systems \& Information Dynamics}
	{\small\bf 24}(3) (2017) 1740002.
	
	\bibitem{gemmer200419}
	J.~Gemmer, M.~Michel, and G.~Mahler,
	{\small\it Quantum thermodynamics: Emergence of thermodynamic behavior within composite quantum systems}
	(Springer-Verlag, Berlin-Heidelberg, 2009).
	
	\bibitem{kosloff2013quantum}
	R.~Kosloff
	Quantum thermodynamics: A dynamical viewpoint,
	{\small\it Entropy}
	{\small\bf 15}(6) (2013) 2100--2128.
	
	\bibitem{trushechkin2016perturbative}
	A.\,S.~Trushechkin and I.\,V.~Volovich
	Perturbative treatment of inter-site couplings in the local description of open quantum networks,
	{\small\it EPL}
	{\small\bf 113}(3) (2016) 30005.
	
	\bibitem{Tru2017}
	A.\,S.~Trushechkin,
	Ob obschem opredelenii proizvodstva entropii v markovskikh otkrytykh kvantovykh sistemakh,
	Kvantovye vychisleniya, Itogi nauki i tekhn. Ser. Sovrem. mat. i ee pril. Temat. obz.
	{\small\it VINITI RAN}
	{\small\bf 138} (2017) 82--98 [in Russian].
	
	\bibitem{DodMan86}
	V.\,V.~Dodonov, V.\,I.~Man'ko,
	Evolution equations for the density matrices of linear open systems,
	in {\small\it Classical and Quantum Effects in Electrodynamics. Proc. Lebedev Phys. Inst.}, Vol. 176 (A.\,A~Komar, ed.), pp. 53--60 (Nova Science, Commack, 1988).
	
	\bibitem{dodonov1985quantum}
	V.\,V.~Dodonov,  O.\,V.~Manko,
	Quantum damped oscillator in a magnetic field,
	{\small\it Physica A}
	{\small\bf 130}(1--2) (1985), 353--366.
	
	\bibitem{sandulescu1987open}
	A.~Sandulescu, H.~Scutaru, W.~Scheid,
	Open quantum system of two coupled harmonic oscillators for application in deep inelastic heavy ion collisions,
	{\small\it J. of Phys. A}
	{\small\bf 20}(8) (1987) 2121.
	
	\bibitem{isar2006damped}
	A.~Isar, A.~Sandulescu,
	Damped quantum harmonic oscillator
	arXiv:quant-ph/0602149 (2006).
	
	\bibitem{isar2013entanglement}
	A.~Isar,
	Entanglement of formation for Gaussian states of two bosonic modes in a thermal environment,
	{\small\it Rom. J. Phys.}
	{\small\bf 58} (2013) 1355.
	
	\bibitem{Prosen10}
	T.~Prosen and T.~H. Seligman,
	Quantization over boson operator spaces,
	{\small\it J. of Phy. A: Math. and Theor.}
	{\small\bf 43}(39) (2010) 392004.
	
	\bibitem{glauber1984damping}
	R.~Glauber and V.\,I.~Man'ko,
	Damping and fluctuations in coupled quantum oscillator systems,
	{\small\it JETP}
	{\small\bf 60} (1984) 450--457.
	
	\bibitem{Holevo15}
	A.\,S.~Holevo,
	{\small\it Quantum systems, channels, information: a mathematical introduction}
	(De~Gruyter, Berlin, 2012).
	
	\bibitem{Parth15}
	K.\,R.~Parthasarathy,
	Symplectic dilations, Gaussian states and Gaussian channels,
	{\small\it Indian Journal of Pure and Applied Mathematics}
	{\small\bf 46}(4) (2015) 419--439.
	
	\bibitem{Grepl13}
	E.~Greplova,
	{\small\it Quantum Information with Fermionic Gaussian States}
	(Munich, 2013).
	
	\bibitem{holevo15gaussopt}
	A.\,S.~Holevo,
	Gaussian optimizers and the additivity problem in quantum information theory,
	{\small\it Russ. Math. Surveys}
	{\small\bf 70}(2) (2015) 331--367.
	
	\bibitem{Hol2016proof}
	A.\,S.~Holevo,
	On the proof of the majorization theorem for quantum Gaussian channels,
	{\small\it Russ. Math. Surveys}
	{\small\bf 71}(3) (2016) 585--587.
	
	\bibitem{Hol2017class}
	A.\,S.~Holevo,
	On the classical capacity of a channel with stationary quantum Gaussian noise,
	{\small\it Theory Probab. Appl.}
	{\small\bf 62}(4) (2018) 534--551.
	
	\bibitem{Hol17}
	A.\,S.~Holevo,
	On the quantum Gaussian optimizers conjecture in the case q=p,
	{\small\it Russ. Math. Surveys}
	{\small\bf 72}(6) (2017) 1177--1179.
	
	\bibitem{Cubitt12}
	T.\,S.~Cubitt, J.~Eisert, M.\,M.~Wolf,
	The complexity of relating quantum channels to master equations,
	{\small\it Comm. in Math. Phys.}
	{\small\bf 310}(2) (2012) 383--418.
	
	\bibitem{heinosaari2009semigroup}
	T.~Heinosaari, A.\,S.~Holevo, M.\,M.~Wolf,
	The semigroup structure of Gaussian channels,
	arXiv:0909.0408 (2009).
	
	\bibitem{Cahill99}
	K.\,E.~Cahill, R.\,J.~Glauber,
	Density operators for fermions,
	{\small\it Phys. Rev. A}
	{\small\bf 59} (1999) 1538.
	
	\bibitem{VlaVol84}
	V.\,S.~Vladimirov, I.\,V.~Volovich,
	Superanalysis. I. Differential calculus,
	{\small\it Theoret. and Math. Phys.}
	{\small\bf 51}(1) (1984) 317--335.
	
	\bibitem{VlaVol84a}
	V.\,S.~Vladimirov, I.\,V.~Volovich,
	Superanalysis. II. Integral calculus,
	{\small\it Theoret. and Math. Phys.}
	{\small\bf 60}(2) (1984) 743--765.
	
	\bibitem{Prosen10a}
	T.~Prosen, B.~Zunkovic,
	Exact solution of Markovian master equations for quadratic Fermi systems: thermal baths, open XY spin chains and non-equilibrium phase transition,
	{\small\it New J. Phys.}
	{\small\bf 12} (2010) 025016.
	
	
	\bibitem{Clarc10}
	S.\,R.~Clark,  J.~Prior, M.\,J.~Hartmann, D.\,Jaksch,  and M.\,B.~Plenio,
	Exact matrix product solutions in the Heisenberg picture of an open quantum spin chain,
	{\small\it New J. Phys.}
	{\small\bf 12}(2) (2010) 025005.
	
	\bibitem{Teretenkov17b}
	A.\,E.~Teretenkov,
	Quadratic Fermionic Dynamics with Dissipation,
	{\small\it Math. Notes}
	{\small\bf 102}(6) (2017) 846--854.
	
	\bibitem{Alicki07}
	R.\,Alicki and K.\,Lendi
	{\small\it Quantum Dynamical Semigroups and Applications}
	(Springer-Verlag, Berlin-Heidelberg, 2004).
	
	\bibitem{Bravyi04}
	S.~Bravyi,
	Lagrangian representation for fermionic linear optics,
	arXiv:quant-ph/0404180 (2004).
	
	\bibitem{knill2001fermionic}
	E.~Knill,
	Fermionic linear optics and matchgates,
	arXiv:quant-ph/0108033 (2001).
	
	\bibitem{terhal2002classical}
	B.\,M.~Terhal, D.\,P.~Di\,Vincenzo,
	Classical simulation of noninteracting-fermion quantum circuits,
	{\small\it Phys. Rev. A}
	{\small\bf 65}(3) (2002) 032325.
	
	\bibitem{kitaev2001unpaired}
	A.\,Yu.~Kitaev,
	Unpaired Majorana fermions in quantum wires,
	{\small\it Physics-Uspekhi}
	{\small\bf 44}(10) (2001) 131--136.
	
	\bibitem{bravyi2002fermionic}
	S.\,B.~Bravyi and A.\,Yu.~Kitaev,
	Fermionic quantum computation
	{\small\it Annals of Physics}
	{\small\bf 298}(1) (2002) 210--226.
	
	\bibitem{Teretenkov16}
	A.\,E.~Teretenkov,
	Quadratic Dissipative Evolution of Gaussian States,
	{\small\it Math. Notes}
	{\small\bf 100}(4) (2016) 642--646.
	
	\bibitem{Teretenkov17}
	A.\,E.~Teretenkov,
	Quadratic Dissipative Evolution of Gaussian States with Drift,
	{\small\it Math. Notes}
	{\small\bf 101}(2) (2017) 341--351.
	
	\bibitem{Chebotarev12}
	A.\,M.~Chebotarev, A.\,E.~Teretenkov,
	Operator-Valued ODEs and Feynman's Formula,
	{\small\it Math. Notes}
	{\small\bf 92}(6) (2012) 837--842.
	
	\bibitem{Teretenkov19}
	A.\,E.~Teretenkov,
	Dynamics of Moments for Quadratic GKSL Generators,
	{\small\it Mat. Zametki}
	{\small\bf 106}(1) (2019) 149--153.
	
	\bibitem{Cheb2000}
	A.~M Chebotarev,
	{\small\it Lectures on quantum probability}
	(Sociedad Matemática Mexicana, México, 2000).
	
	\bibitem{Holevo2003}
	A.\,S.~Holevo,
	{\small\it Statistical structure of quantum theory}
	(Springer-Verlag, Berlin-Heidelberg, 2003).
	
	\bibitem{colpa1978diagonalization}
	J.\,H.\,P.~Colpa,
	Diagonalization of the quadratic boson Hamiltonian,
	{\small\it Phys. A: Stat. Mech. and its Appl.}
	{\small\bf 93}(3--4) (1978) 327--353.
	
	\bibitem{Tak11}
	L.\,A.~Takhtajan,
	{\small\it Quantum Mechanics for Mathematicians}
	(American Mathematical Society, Providence, 2008).
	
	\bibitem{carmichael2000statistical}
	H.\,J.~Carmichael and M.\,O.~Scully,
	{\small\it Statistical methods in quantum optics 1: master equations and Fokker-Planck equations}
	(Springer-Verlag, Berlin-Heidelberg, 2013).
	
	\bibitem{combescure2012coherent}
	M.~Combescure, D.~Robert,
	{\small\it Coherent states and applications in mathematical physics}
	(Springer-Verlag, Berlin-Heidelberg, 2012).
	
	\bibitem{Amosov17}
	G.\,G.~Amosov and S.\,N.~Filippov,
	Spectral properties of reduced fermionic density operators and parity superselection rule
	{\small\it Quant. Inform. Proc.}
	{\small\bf 16}(1) (2017) 2.
	
	\bibitem{Holevo18}
	A.\,S.~Holevo,
	On Singular Perturbations of Quantum Dynamical Semigroups
	{\small\it Math. Notes}
	{\small\bf 103}(1--2) (2018) 133--4144.
	
	\bibitem{quadratica2017}
	https://github.com/LeXMSU/Quadratica
	
	\bibitem{maniscalco2006non}
	S.~Maniscalco and F.~Petruccione,
	Non-Markovian dynamics of a qubit,
	{\small\it Phys. Rev. A}
	{\small\bf 73}(1) (2006) 012111.
	
	\bibitem{fleming2012non}
	C.\,H.~Fleming and B.\,L.~Hu,
	Non-Markovian dynamics of open quantum systems: stochastic equations and their perturbative solutions,
	{\small\it Annals of Physics}
	{\small\bf 327}(4) (2012) 1238--1276.
	
	\bibitem{de2017dynamics}
	I.~De\,Vega and D.\,Alonso,
	Dynamics of non-Markovian open quantum systems,
	{\small\it Rev. of Mod. Phys.}
	{\small\bf 89}(1) (2017) 015001.
	
	\bibitem{semin2016minimalistic}
	V.~Semin and F.~Petruccione,
	Minimalistic analytical approach to non-Markovian open quantum systems,
	{\small\it EPL}
	{\small\bf 113}(2) (2016) 20004.
	
	\bibitem{breuer2001time}
	H.-P.~Breuer, B.~Kappler, and F.~Petruccione,
	The time-convolutionless projection operator technique in the quantum theory of dissipation and decoherence,
	{\small\it Annals of Physics}
	{\small\bf 291}(1) (2001) 36--70.
	
	\bibitem{maniscalco2004lindblad}
	S.~Maniscalco, J.~Piilo, F.~Intravaia, F.~Petruccione, and A.~Messina,
	Lindblad-and non-Lindblad-type dynamics of a quantum Brownian particle,
	{\small\it Phys. Rev. A}
	{\small\bf 70}(3) (2004) 032113.
	
	\bibitem{maniscalco2009non}
	S.~Maniscalco, J.~Piilo, and K.\,A.~Suominen,
	Non-Markovian weak coupling limit of quantum Brownian motion,
	{\small\it The European Phys. J. D}
	{\small\bf 55}(1) (2009) 181.
	
	\bibitem{shaji2005s}
	A.~Shaji and E.\,C.\,G.~Sudarshan,
	Who's afraid of not completely positive maps?
	{\small\it Phys. Lett. A}
	{\small\bf 341}(1--4) (2005) 48--54.
	
	\bibitem{man2005partial}
	O.\,V.~Man'ko, V.\,I.~Man'ko, G.~Marmo, A.~Shaji, E.\,C.\,G.~Sudarshan, and F.~Zaccaria,
	Partial positive scaling transform: a separability criterion,
	{\small\it Phys. Lett. A}
	{\small\bf 339}(3--5) (2005) 194--206.
	
	\bibitem{DePalma15}
	G.~De\,Palma, A.\,Mari, V.\,Giovannetti, and A.\,S.~Holevo,
	Normal form decomposition for Gaussian-to-Gaussian superoperators,
	{\small\it J. of Math. Phys.}
	{\small\bf 56}(5) (2015) 052202.
	
	\bibitem{Dumcke85} D\"umcke, R.:
	The low density limit for an N-level system interactig with a free Bose or Fermi Gas.
	{\small\it Comm. Math. Phys.} {\small\bf 97} (1985) 331-359
	
	\bibitem{filippov2012local}
	S.\,N.~Filippov, T.~Rybár, and M.~Ziman,
	Local two-qubit entanglement-annihilating channels,
	{\small\it Phys. Rev. A}
	{\small\bf 85}(1) (2012) 012303.
	
	\bibitem{maniscalco2007entanglement}
	S.~Maniscalco, S.~Olivares, and M.\,G.\,A.~Paris,
	Entanglement oscillations in non-Markovian quantum channels,
	{\small\it Phys. Rev. A}
	{\small\bf 75}(6) (2007) 062119.
	
	\bibitem{chruscinski2010memory}
	D.~Chruscinski and J.~Jurkowski,
	Memory in a nonlocally damped oscillator, {\small\it in Quantum Bio-Informatics III} (2010), pp. 155--166.
	
	\bibitem{chruscinski2010non}
	D.~Chruscinski and A.~Kossakowski,
	Non-Markovian quantum dynamics: local versus nonlocal,
	{\small\it Phys. Rev. Lett.}
	{\small\bf 104}(7) (2010) 070406.
	
	\bibitem{chruscinski2014time}
	D.~Chruscinski,
	On time-local generators of quantum evolution,
	{\small\it Open Systems \& Information Dynamics}
	{\small\bf 21}(01--02) (2014) 1440004.
	
	\bibitem{siudzinska2017memory}
	K.~Siudzinska and D.~Chruscinski,
	Memory kernel approach to generalized Pauli channels: Markovian, semi-Markov, and beyond,
	{\small\it Phys. Rev. A}
	{\small\bf 96}(2) (2017) 022129.
	
	\bibitem{Holevo93}
	A.\,S.~Holevo,
	A note on covariant dynamical semigroups,
	{\small\it Rep. on Math. Phys.}
	{\small\bf 32}(2) (1993) 211--216.
	
	\bibitem{Holevo96}
	A.\,S.~Holevo,
	Covariant quantum Markovian evolutions,
	{\small\it J. of Math. Phys.}
	{\small\bf 37}(4) (1996) 1812--1832.
	
	\bibitem{Holevo95}
	A.\,S.~Holevo,
	On the Structure of Covariant Dynamical Semigroups,
	{\small\it J. of Funct. Anal.}
	{\small\bf 131}(2) (1995) 255--278.
	
\end{thebibliography}
\end{document}